\documentclass[11pt]{article}
\usepackage[margin=1in]{geometry}
\usepackage{times}
\usepackage{microtype}
\usepackage{graphicx}
\usepackage{amsmath, amssymb}
\usepackage{booktabs}
\usepackage{hyperref}
\usepackage{enumitem}
\usepackage{xcolor}
\usepackage{subcaption}
\usepackage{array}
\usepackage{ragged2e}

\graphicspath{{figures/}}

\newcolumntype{L}[1]{>{\RaggedRight\arraybackslash}p{#1}}

\title{\textbf{Nova3D}: Code-Native Generation of Programmable 3D Assets\\
\large Generating structured, editable, constraint-consistent 3D assets as
executable programs}
\author{
Nimra Noor$^{1}$ \quad Muhammad Bilal$^{1,2}$ \quad Abdullah Hussain$^{1}$ \quad Hassan Baig$^{1}$\\[4pt]
$^{1}$Raresense Inc \qquad $^{2}$University of Edinburgh\\[6pt]
\normalsize Code and benchmark: \url{https://github.com/RareSense/Nova3D}
}
\date{}

\begin{document}
\maketitle

\begin{abstract}
Current 3D generative models mostly produce a final surface: a visually strong
but largely opaque mesh. Interactive 3D worlds need more than a surface. They
need named parts, an assembly hierarchy, measurable constraints, local edit
handles, and joints for articulation. We present \textbf{Nova3D}, a system that
generates 3D assets as executable Blender source code; the compiled mesh, a
binary glTF (GLB), is treated as the artifact, not the asset. Because the output
is a program, semantic handles exist at generation time rather than being
recovered afterward by segmentation or rigging. We evaluate on
\textbf{Nova3D-Bench}, a frozen, spec-grounded benchmark of 54 items across six
domains and three difficulty levels with text and image inputs, against eleven
baselines in four families (mesh-native, part-structured, code-native, and CAD)
plus a same-LLM ablation. Nova3D produces an executable program and a valid
artifact for 54/54 items. Every asset exposes named parts organized in a
parent--child assembly tree; no mesh-native, CAD, or segmentation baseline
exposes either. It satisfies 51/52 prompt-stated numeric and count constraints
(best baseline: 11/52), passes 14/18 blinded local edits with locality preserved
in 18/18, and articulates 59 joints across 12 assets at 98.3\% geometric
validity, where every baseline exposes zero native joints. Its geometry is
competitive: it wins the structured domains in a pairwise shape-quality
tournament and is second only to the strongest mesh-native model, while
conceding texture realism to baked-PBR systems. The central result is
representational: code-native generation turns a generated 3D object from an
opaque surface into a programmable asset that downstream systems can inspect,
measure, edit, and animate.
\end{abstract}

\section{Introduction}
Mesh-native 3D generation has become remarkably good at synthesizing
surfaces~\cite{poole2022dreamfusion, lin2023magic3d, xiang2024trellis,
li2025triposg, zhao2025hunyuan3d}. Given a prompt, these systems produce a
visually convincing mesh. But a generated object that will live in an
interactive world is not just something to render; it is something an agent,
script, physics engine, or animator must act on. Acting on an object requires
handles: the ability to name a part, count it, measure it, move it, edit it, and
attach a joint to it. A mesh-native generator delivers a fused surface with, at
best, a few generic nodes (\texttt{geometry\_0}). Those handles do not come for
free; they have to be recovered afterward, through a separate segmentation,
naming, and rigging pipeline.

\begin{figure}[ht!]
\centering
\includegraphics[width=\linewidth]{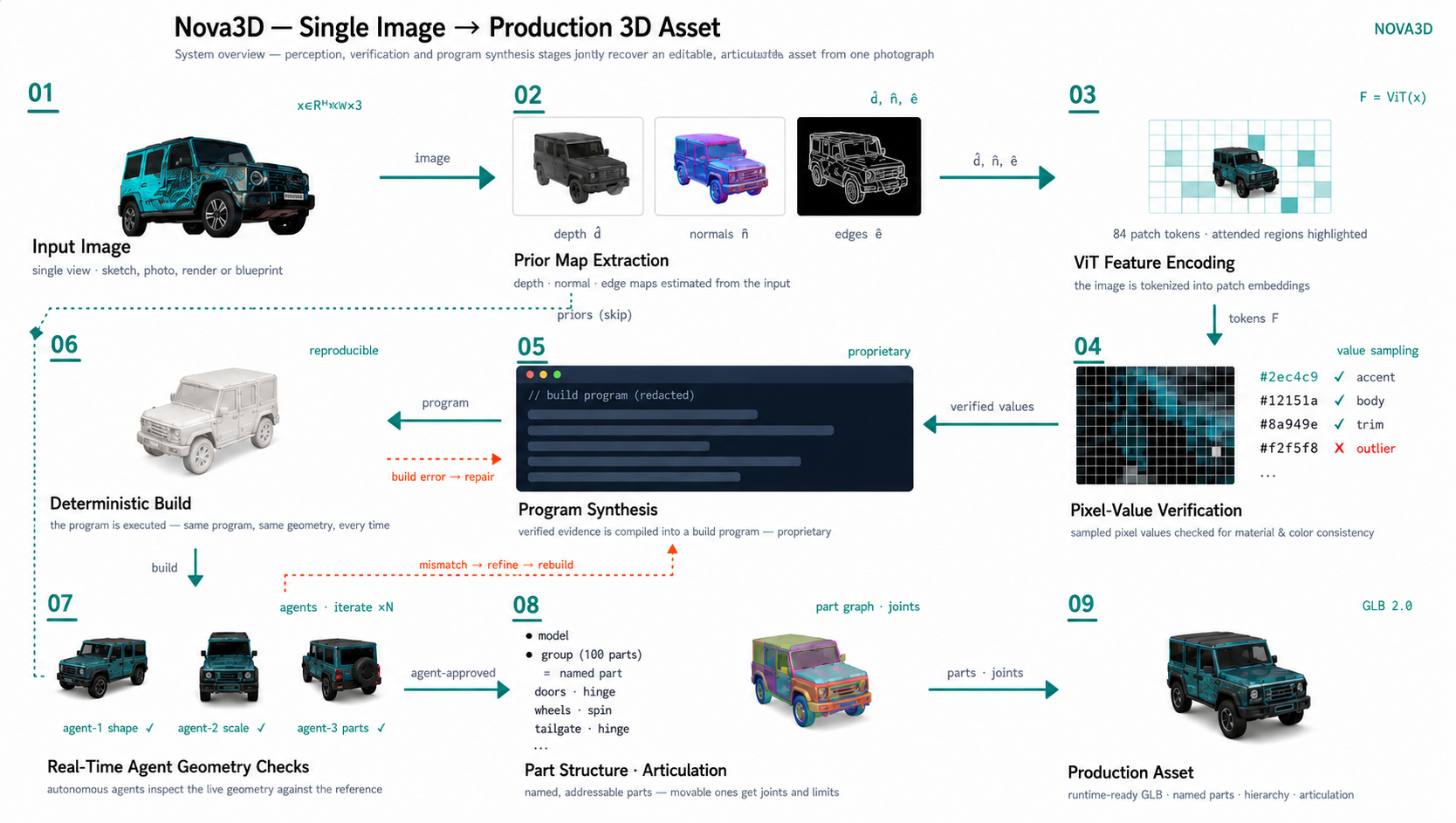}
\caption{\textbf{Nova3D system overview: single input to production 3D asset.}
From one input view (sketch, photo, render, or blueprint), the system extracts
prior maps (depth, surface normals, edges) and ViT patch features (stages
01--03), verifies sampled pixel values for material and color consistency,
rejecting outliers (04), and compiles the verified evidence into an executable
build program (05). The program is executed deterministically in headless
Blender, with build errors routed back for repair (06); autonomous agents then
inspect the live geometry against the reference for shape, scale, and parts,
and any mismatch triggers a refine-and-rebuild cycle (07). The approved asset
is finalized as a part graph of named, addressable parts whose movable ones
carry joints and limits (08), and exported as a runtime-ready GLB with names,
hierarchy, and articulation (09). Text-conditioned items enter directly at
program synthesis.}
\label{fig:architecture}
\end{figure}

\begin{figure}[ht!]
\centering
\includegraphics[width=0.92\linewidth]{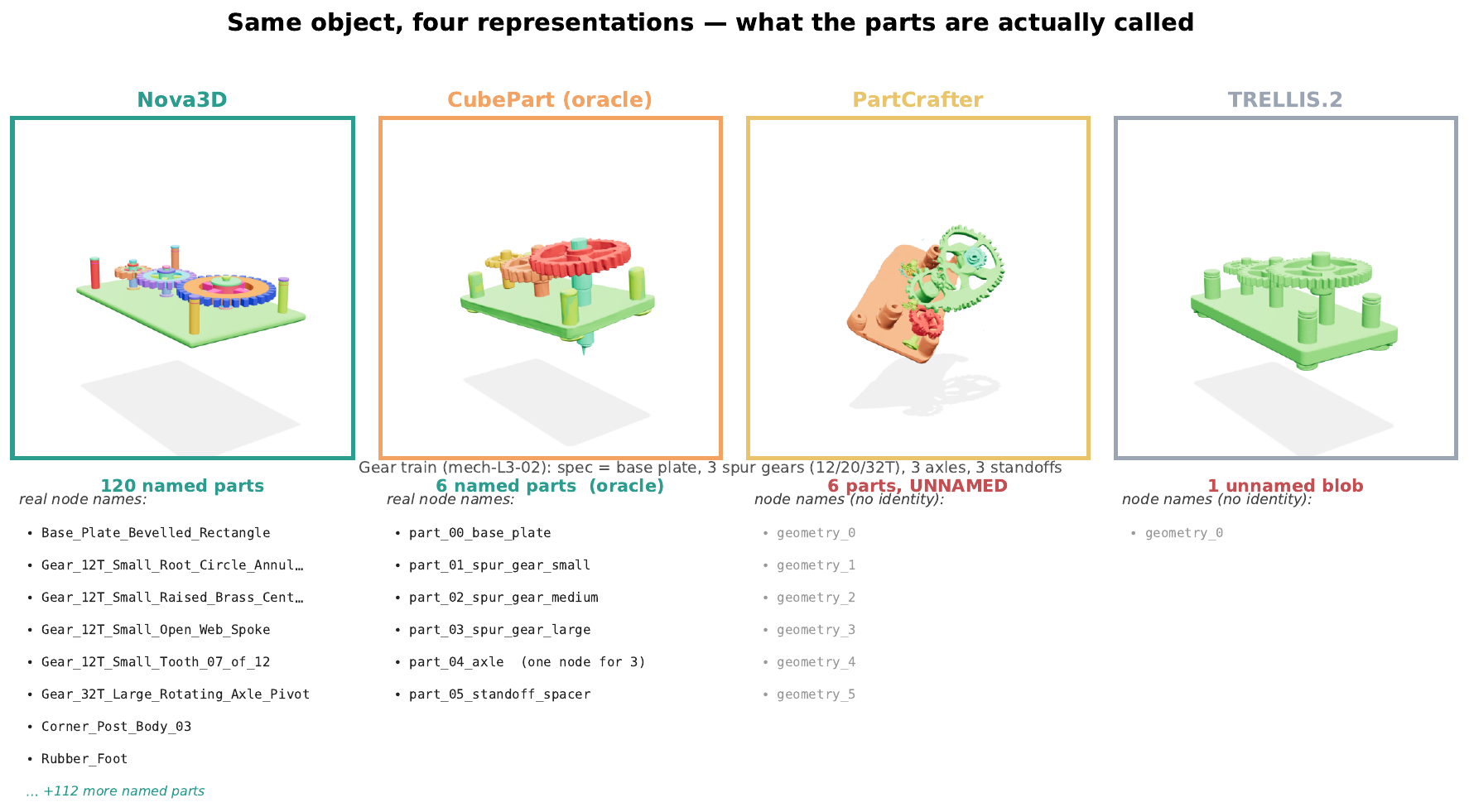}
\caption{\textbf{The program is the asset: native semantic handles, shown rather
than told.} A generated gear train rendered alongside each system's \emph{actual}
exported node names. Nova3D exposes individually named teeth, hubs, and pivots
organized into a parent--child tree. CubePart, the oracle post-hoc segmenter that
operates on a TripoSG mesh and is handed the target part names, returns a short
flat list of labels (capped); the part-aware diffusion model returns unnamed
\texttt{geometry\_N}; the mesh-native model returns a single blob. These handles
are present at generation time in Nova3D, and must be recovered by post-hoc
segmentation and rigging for the others.}
\label{fig:namingexample}
\end{figure}

Nova3D takes a different route. Instead of generating the final surface, it
generates the asset's \emph{source code}: a self-contained Blender Python
program whose compiled GLB is merely the output of running it. The guiding
principle is simple. \textbf{The program is the asset; the mesh is a compiled
artifact.} Because the asset is a program, the structure an interactive world
needs is present at generation time. Where a mesh-native generator emits
\texttt{geometry\_0}, Nova3D emits \texttt{Gear\_12T\_Small\_Tooth\_07} and
\texttt{Gear\_32T\_Large\_Rotating\_Axle\_Pivot}, organized into a parent--child
hierarchy with pivots, materials, and edit handles
(Figure~\ref{fig:namingexample}). Figure~\ref{fig:architecture} gives the system
overview.

This is a representation shift, not a claim of visual superiority. Nova3D does
not aim to beat mesh-native generators on texture realism or freeform organic
surfaces. Those methods are strong there, and our results place Nova3D close
behind the best of them, which anchors the comparison rather than overturning
it. The contribution is that an executable program keeps the asset inspectable,
addressable, measurable, editable, and animatable, while a baked mesh does not
natively expose any of these without post-hoc processing. We summarize the
thesis as:

\begin{quote}
\emph{Nova3D shows that generating 3D as code preserves parity-or-better shape
quality in structured domains while adding structural affordances that
mesh-native generators do not natively expose.}
\end{quote}

We support this with a single causal chain rather than a catalog of metrics.
(1)~Nova3D \emph{reliably} generates executable Blender source and a valid GLB,
and a same-model ablation without the Nova3D system package does not, which
answers the natural objection that this is ``just prompting an LLM to write
Blender.'' (2)~The resulting assets are visually and geometrically
\emph{adequate}: competitive on shape, conceding texture. (3)~The code
representation exposes \emph{native semantic structure}: named parts and a real
assembly hierarchy. (4--6)~That structure is what makes constraints
\emph{measurable}, local edits \emph{possible}, and articulation
\emph{possible}. Throughout, the comparison is not a single universal
leaderboard; each axis uses the baseline families for which that representation
exposes the relevant affordance, stated explicitly in Table~\ref{tab:coverage}.

\paragraph{Contributions.}
\begin{enumerate}[leftmargin=2em, labelsep=0.5em]
    \item \textbf{Nova3D}, a code-native 3D generation pipeline that turns text,
    image, or sketch inputs into an executable Blender program and a compiled GLB
    with named parts, hierarchy, pivots, and materials, refined by a closed-loop
    system package combining perception, verification, execution, and render
    feedback.
    \item \textbf{Nova3D-Bench}, a frozen, spec-grounded benchmark of 54 items for
    evaluating \emph{programmable} 3D assets, with machine-checkable ground truth
    for parts, counts, joints, dimensions, and symmetry.
    \item A comprehensive evaluation spanning reliability, visual/geometry quality,
    semantic structure, constraints, editing, and articulation, against four
    baseline families under an input-parity rule.
    \item Evidence that code-native assets expose programmable structure, and
    support downstream constraint measurement, local editing, and articulation,
    that mesh-native outputs do not natively expose.
\end{enumerate}

\section{Related Work}

The dominant line of work in generative 3D synthesizes the final surface
directly. Optimization-based systems such as
DreamFusion~\cite{poole2022dreamfusion} and Magic3D~\cite{lin2023magic3d}
distill 2D diffusion priors into 3D representations, and feed-forward systems
including Point-E~\cite{nichol2022pointe}, Shap-E~\cite{jun2023shape},
Zero-1-to-3~\cite{liu2023zero123}, TripoSG~\cite{li2025triposg},
TRELLIS~\cite{xiang2024trellis}, Hunyuan3D~\cite{zhao2025hunyuan3d}, and the
commercial Meshy~\cite{meshy2024} have steadily improved speed, fidelity, and
texture quality. These methods are the right comparison class for perceptual
fidelity, and we treat them as such. Their limitation is representational
rather than visual: the output is a fused mesh or an opaque reconstruction with
a few generic nodes, so addressing ``front wheel'' or ``elbow pivot'' requires
a separate segmentation and rigging stage after generation. Nova3D is
complementary to this line of work in that it makes the construction procedure
and the scene graph part of the output itself.

A second family targets part structure. Part123~\cite{liu2024part123},
PartGen~\cite{chen2024partgen}, and PartCrafter~\cite{lin2025partcrafter}
recover or jointly generate part-level meshes from image cues, while
segmentation methods such as PartSLIP~\cite{liu2023partslip},
SAMPart3D~\cite{yang2024sampart3d}, and Find3D~\cite{ma2024find3d} decompose a
given shape, with PartNet~\cite{mo2019partnet} and the Name-That-Part
benchmark~\cite{paul2025namethatpart} providing the standard datasets and
label-aware metrics. The outputs of this family are typically flat or
schema-conditioned: unnamed part meshes, or region labels that require the
answer vocabulary as input. Our evaluation therefore includes PartCrafter as a
closed-book born-segmented baseline and CubePart as an oracle post-hoc
segmenter that operates on TripoSG meshes and is handed the spec part names,
and it scores native named handles and assembly structure rather than
segmentation IoU, for which the benchmark has no shared point-level masks.

Closest to our approach are systems in which a language model writes the 3D
content as code. 3D-GPT~\cite{sun20243dgpt}, L3GO~\cite{yamada2024l3go},
SceneCraft~\cite{hu2024scenecraft}, LL3M~\cite{ll3m2024}, and
BlenderLLM~\cite{du2024blenderllm} emit procedural or Blender scripts and
validate the premise that code is a natural medium for structured 3D. They also
expose the practical gap Nova3D targets: several are scene-layout rather than
single-asset generators, some depend on unreleased backends, and vanilla script
generation lacks robust repair, validation, hierarchy discipline, and reliable
export. The CAD literature pursues the same representational idea in a narrower
substrate: DeepCAD~\cite{wu2021deepcad}, Text2CAD~\cite{khan2024text2cad},
Text-to-CadQuery~\cite{texttocadquery2025}, CAD-Coder~\cite{cadcoder2025}, and
CADFusion~\cite{cadfusion2025} produce editable CAD programs or B-rep solids
that are valid, compact, and often watertight, but that usually compile to a
fused solid or a few generic bodies and are strongest on mechanical prompts
rather than textured multi-domain assets. We compare directly against
BlenderLLM, against Text2CAD, Text-to-CadQuery, and CAD-Coder as runnable CAD
baselines, and against a naive same-LLM ablation that uses the identical model
and route without the Nova3D system package, isolating the contribution of the
system rather than the base model.

Finally, our evaluation protocol builds on prior work in automatic 3D
assessment. GPTEval3D~\cite{wu2024gpteval3d} showed that GPT-4V-style pairwise
judgments over rendered views align with human preference and aggregate into
Elo rankings, and we follow this protocol for perceptual fidelity while keeping
the visual track separate from structural claims. Executable CAD benchmarks
such as CADSmith~\cite{barkley2026cadsmith} and
CADTestBench~\cite{zhang2026cadtestbench} motivate our choice to score
constraints with automated tests rather than visual scores alone. Existing
benchmarks each target a different slice of the problem: T3Bench~\cite{he2023t3bench}
and GT23D-Bench~\cite{su2024gt23dbench} measure rendered quality and text
alignment, Objaverse~\cite{deitke2023objaverse} provides scale without
controlled construction tasks, and ABC~\cite{koch2019abc},
Fusion~360~\cite{willis2021fusion360}, and GenCAD-Code~\cite{li2025gencadcode}
are tied to parametric CAD substrates. Nova3D-Bench addresses the intersection
these miss and is a complement to them, not a replacement.

\section{Method: Nova3D Code-Native Generation}
Nova3D's contribution is a representation contract, not a specific prompt. The
asset is generated as a program that, when executed, builds a named,
hierarchical, constraint-checkable object (Figure~\ref{fig:architecture}).

\subsection{Inputs and outputs}
Nova3D accepts a text prompt, a sketch or reference image, or an existing design
paired with an edit instruction. Each run produces an executable
\texttt{code.py} Blender program, a compiled \texttt{model.glb}, named mesh
nodes, named group nodes, joint/pivot handles, and materials with vertex-baked
ambient occlusion. The runtime, prompt scaffolding, and orchestration are fixed
across all six evaluated domains; only the per-item request varies. Generation
sees only the manifest prompt (and reference image where present), never the
benchmark spec, so the naming and structure results in
Sections~\ref{sec:structure}--\ref{sec:articulation} are closed-book.

\subsection{Code as the source of truth}
\begin{figure}[ht!]
\centering
\begin{subfigure}{0.45\linewidth}
\includegraphics[width=\linewidth]{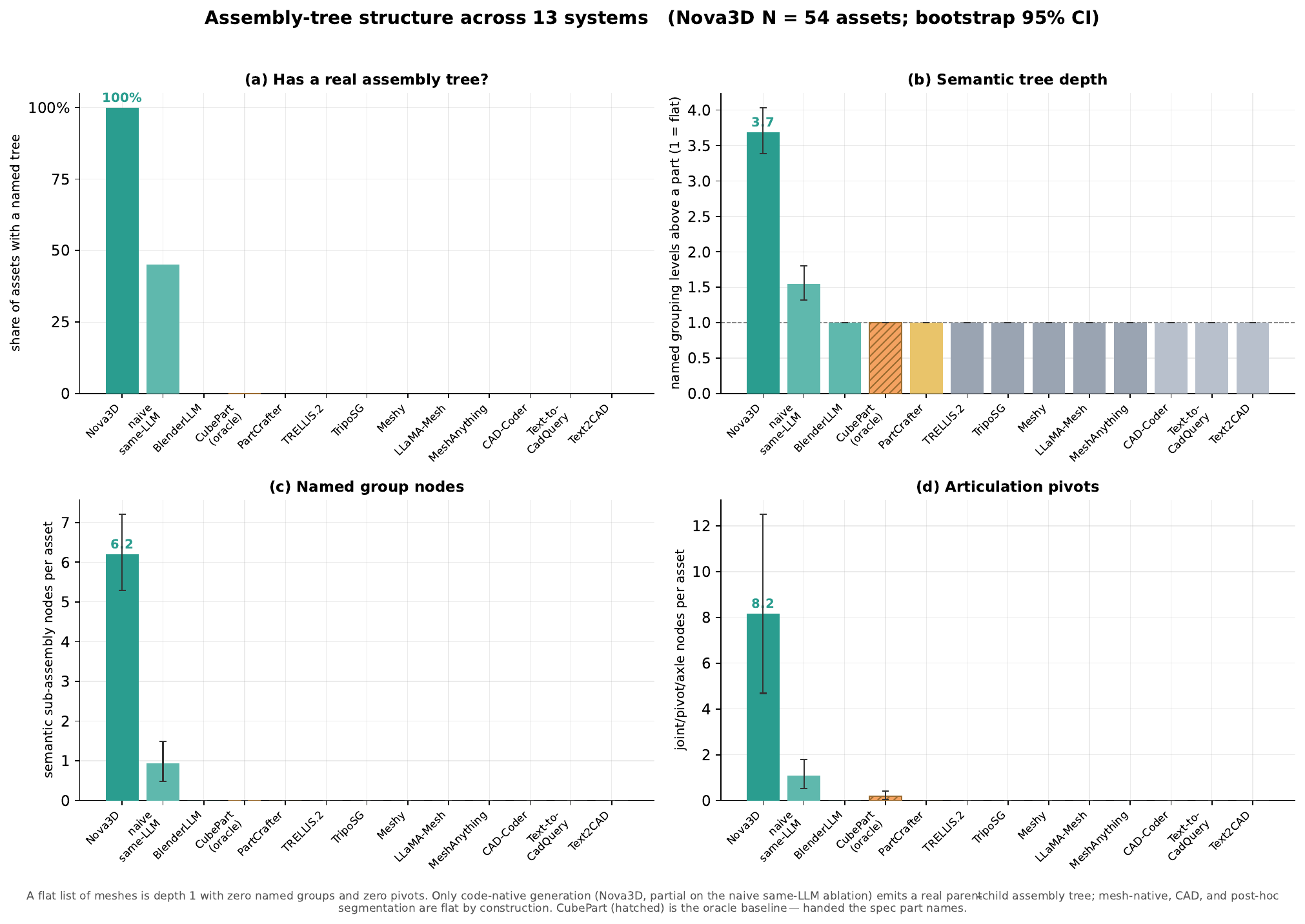}
\caption{Assembly-tree statistics}
\end{subfigure}\hfill
\begin{subfigure}{0.53\linewidth}
\includegraphics[width=\linewidth]{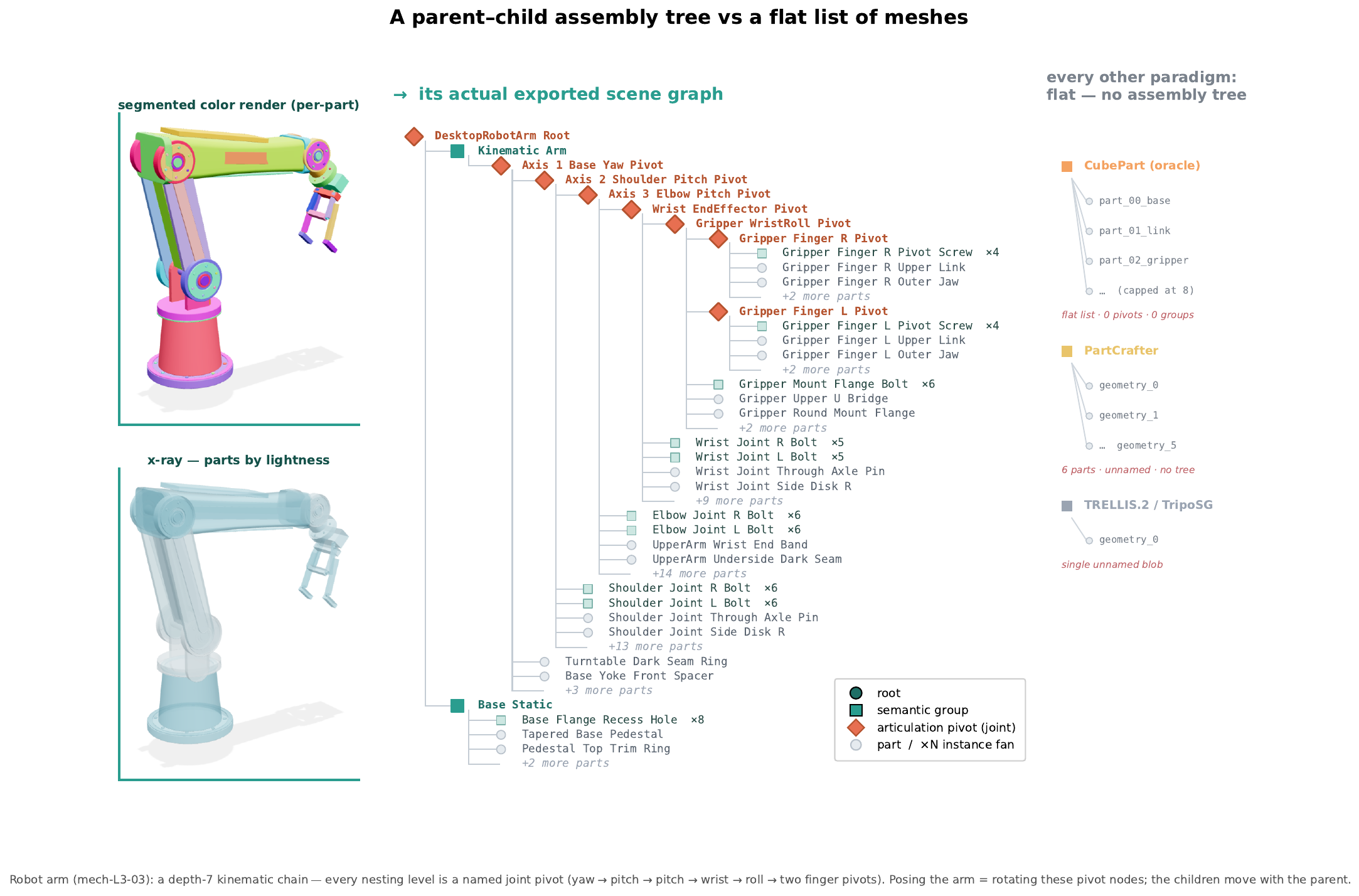}
\caption{A tree vs a flat list (robot arm)}
\end{subfigure}
\caption{\textbf{A parent--child assembly tree is exposed natively only by
code-native generation.} The robot arm's actual scene graph is a depth-7 kinematic
chain, every level a named pivot; the baselines are flat lists with nothing to
address or move.}
\label{fig:hierarchy}
\end{figure}

The Blender source stores construction intent; the GLB is the compiled result.
Named objects in the source become handles in the exported scene graph, and
those handles are what later make constraints measurable, edits local, and
articulation possible. This is the property that post-hoc segmentation of a
finished mesh cannot recover: the construction logic was never present to begin
with. Figure~\ref{fig:code} shows an excerpt of an actual generated program; the
same asset's exported scene graph appears in Figure~\ref{fig:hierarchy}b.

\begin{figure}[ht!]
\centering
\begin{minipage}{0.9\linewidth}
\small
\begin{verbatim}
# Dimensions are named constants derived from the request
BASE_FLANGE_R   = 0.170
BASE_BOLT_COUNT = 8
SHOULDER = Vector((0.0,  0.000, 0.405))   # physical joint locations
ELBOW    = Vector((0.0, -0.105, 0.865))

def set_pivot(obj, pivot_world):          # part origin = its pivot
    obj.data.transform(Matrix.Translation(-Vector(pivot_world)))
    obj.location = Vector(pivot_world)

core = build_arm_core("Lower_Arm_Core")   # one named mesh per component
set_pivot(core, SHOULDER)                 # rotating this node IS the motion
parent_to(core, shoulder_parent)          # under Axis_2_Shoulder_Pitch
\end{verbatim}
\end{minipage}
\caption{\textbf{Excerpt from a generated program} (the desktop robot arm,
abridged). Dimensions are named constants, each component is a separately named
mesh, its origin is set at the physical pivot, and it is parented into a named
joint group. This is why a dimension is checkable against the source's own
anchors and why ``make it taller'' is a one-constant edit.}
\label{fig:code}
\end{figure}

\subsection{Perception, verification, and program synthesis}
\label{sec:pipeline}
The pipeline in Figure~\ref{fig:architecture} recovers an editable asset from a
single conditioning input through a sequence of perception, verification,
synthesis, and validation stages. Let $x \in \mathbb{R}^{H \times W \times 3}$
denote the input view (a sketch, photograph, render, or blueprint) and $t$ the
textual request; text-conditioned items omit $x$ and enter the pipeline directly
at program synthesis. From $x$, the system first estimates a set of geometric
prior maps,
\begin{equation}
(\hat{d},\, \hat{n},\, \hat{e}) \;=\; \Phi(x),
\label{eq:priors}
\end{equation}
where $\hat{d}$, $\hat{n}$, and $\hat{e}$ are monocular depth, surface-normal,
and edge estimates that expose the object's coarse geometry, orientation, and
silhouette independently of its texturing. In parallel, the image is tokenized
into patch embeddings by a vision transformer, $F = \mathrm{ViT}(x) \in
\mathbb{R}^{N \times d}$, whose attended regions indicate which parts of the
view carry structural evidence.

Appearance evidence is verified before any code is written. The system draws a
sample set $S = \{(u_i, c_i)\}_{i=1}^{m}$ of image locations $u_i$ and their
observed colors $c_i$, aggregates them into candidate material assignments
(body, trim, accent), and retains only those whose consistency score exceeds a
threshold,
\begin{equation}
\mathcal{C}^{*} \;=\; \{\, c \;:\; \rho(c \mid S) \ge \tau \,\},
\label{eq:verify}
\end{equation}
rejecting outliers. Only verified values propagate forward, so the program's
material choices are grounded in observed evidence rather than guessed.
Program synthesis then compiles this evidence into an executable build program.
Conditioned on the request, the priors, the attended features, the verified
palette, and the representation doctrine $\mathcal{D}$ of
Section~\ref{sec:doctrine}, a frontier language model $\mathcal{M}_\theta$
emits the initial program
\begin{equation}
P_0 \;=\; \mathcal{M}_\theta\!\left(t,\, \hat{d},\, \hat{n},\, \hat{e},\, F,\,
\mathcal{C}^{*};\, \mathcal{D}\right),
\label{eq:synthesis}
\end{equation}
a self-contained Blender Python script that constructs the object part by part.

Execution grounds the synthesized program in geometry. Headless Blender
implements a deterministic operator
\begin{equation}
\mathrm{Exec}(P_i) \;=\; (G_i,\, R_i,\, \varepsilon_i),
\label{eq:exec}
\end{equation}
mapping a program $P_i$ to its scene graph and geometry $G_i$, a fixed set of
canonical renders $R_i$, and execution diagnostics $\varepsilon_i$ (error
traces, offending API calls, partial scene state); determinism guarantees that
identical programs yield identical geometry, making every downstream
measurement reproducible. Execution acts as a ground-truth validator for
syntax, API compatibility, and basic constructibility, failure modes the
language model cannot reliably detect from source alone. After a successful
build, a set of autonomous checker agents $\{\mathcal{A}_k\}$ inspects the live
geometry against the reference along complementary aspects (shape, scale, and
part coverage in our configuration), producing verdicts $v_k =
\mathcal{A}_k(R_i, x)$. Any build error or agent-detected mismatch triggers a
repair iteration in which the model revises the program conditioned on the
accumulated evidence,
\begin{equation}
P_{i+1} \;=\; \mathcal{M}_\theta\!\left(P_i,\, \varepsilon_i,\,
\{v_k\};\, \mathcal{D}\right), \qquad i < i_{\max} = 3,
\label{eq:repair}
\end{equation}
and the revision is re-executed under the same protocol. The loop terminates
when all agents approve the asset, or when the retry budget is exhausted, in
which case a graceful fallback returns the best successful intermediate
artifact. The approved program is finalized as a part graph of named,
addressable components, movable parts are assigned joints and limits, and the
asset is exported as a runtime-ready GLB carrying names, hierarchy, and
articulation.

Our experiments isolate this \emph{system package} as a whole, comprising
perception (Eq.~\ref{eq:priors}), verification (Eq.~\ref{eq:verify}),
doctrine-conditioned synthesis (Eq.~\ref{eq:synthesis}), execution feedback,
agent checks, and repair
(Eqs.~\ref{eq:exec}--\ref{eq:repair}), against the same base model without it
(Section~\ref{sec:reliability}). We do not claim that any single component,
such as the repair loop alone, is independently responsible for the gains.

\subsection{Representation doctrine}
\label{sec:doctrine}
The export requirements encode a doctrine that makes the asset programmable.
Every visually distinct component is a separate, semantically named mesh, and
repeated instances are named individually (\texttt{Spoke\_17}, never one merged
\texttt{Spokes}). Meshes are parented into a real transform tree with one root
node and semantic group nodes (\texttt{Body}, \texttt{Doors},
\texttt{Left\_Arm}) that exports directly as glTF scene-graph structure. Every
part that should move owns its origin at its physical pivot: a door's hinge
edge, a wheel's axle, a limb's proximal joint, so rotating that node is the
correct motion. Dimensions are named constants derived from the request and its
stated constraints rather than scattered literals, which is what later makes a
dimension checkable against the source's own anchors and makes ``make it
taller'' a one-constant edit. Materials are a small set of physically based
families assigned by part meaning.

The doctrine holds because it is enforced, not merely requested. After a
successful build, canonical renders are compared against the request by a
vision-capable LLM, which either accepts the asset or proposes a corrected
program; a deterministic guard then rejects any correction that shrinks the
source by more than 15\%, drops hierarchy constructs, or loses the entry point,
in which case the pipeline falls back to the last successful artifact. The same
contract governs downstream operations: local edits and articulation are
additive, surgical changes to the source. Articulation adds joint nodes at
pivots and re-parents existing meshes without moving a vertex, which is why
edits preserve non-target content (Section~\ref{sec:editability}) and the rest
pose stays frozen under articulation (Section~\ref{sec:articulation}). The
remainder of the paper measures the affordances this doctrine produces.

\section{Benchmark and Baselines}

\subsection{Nova3D-Bench}
\label{sec:bench}
Nova3D-Bench is a frozen benchmark of \textbf{54 items} organized as
\textbf{6 domains $\times$ 3 difficulty levels $\times$ 3 items}
(Table~\ref{tab:dataset}). Domains are chosen as distinct \emph{construction
strategies} rather than visual categories. Difficulty is defined from the spec,
not prompt length: L1 has $\le 8$ parts and no joints; L2 has 9--25 parts and 1--2
joints; L3 has $>25$ parts or $\ge 3$ joints. Each domain contributes 4 text- and
5 image-conditioned items (24 text, 30 image overall), and one item per
domain$\times$level cell (18 total) carries a multi-constraint set covering
lengths, exact counts, angles, and gear module/pitch diameters, stated in the
prompt text the pipeline sees.

\begin{figure}[ht!]
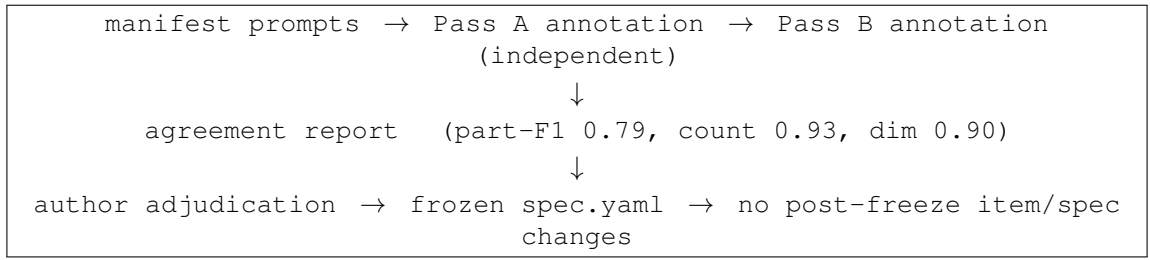

\centering
\fbox{\begin{minipage}{0.9\linewidth}\ttfamily\small
\centering
manifest prompts \;$\rightarrow$\; Pass A annotation \;$\rightarrow$\; Pass B annotation (independent)\\[2pt]
$\downarrow$\\[2pt]
agreement report \; (part-F1 0.79, count 0.93, dim 0.90)\\[2pt]
$\downarrow$\\[2pt]
author adjudication \;$\rightarrow$\; frozen spec.yaml \;$\rightarrow$\; no post-freeze item/spec changes
\end{minipage}}
\caption{\textbf{Benchmark freeze protocol.} A dual-pass, AI-assisted,
author-adjudicated annotation pipeline with a reported agreement gate. There is no
independent human annotation; the protocol and agreement numbers make the ground
truth auditable.}
\label{fig:freeze}
\end{figure}

Each item ships a machine-checkable \texttt{spec.yaml} ground truth (parts and
counts, expected joints, symmetry, key dimensions with explicit \texttt{measure:}
recipes). Ground truth was produced by a dual-pass, AI-assisted, author-adjudicated
protocol (Figure~\ref{fig:freeze}): an independent Pass A and Pass B (different
models, fresh context) were checked for agreement (\textbf{part-F1 0.79},
\textbf{count agreement 0.93}, \textbf{dimension agreement 0.90}, mean), then
author-adjudicated to a canonical functional-part level and frozen. We explicitly
\textbf{do not claim independent human annotation}; multi-pass cross-model
agreement plus author adjudication is the auditability mechanism, and after freeze
no item or spec changes (failures remain in the results).

\begin{table}[ht!]
\centering
\caption{\textbf{Nova3D-Bench composition.} Six construction-strategy domains,
nine items each (three per difficulty level), 4 text / 5 image, 3 constrained
(one per level).}
\label{tab:dataset}
\begin{tabular}{lcccl}
\toprule
Domain & L1/L2/L3 & Text/Image & Constrained & Representative items \\
\midrule
Furniture \& household & 3/3/3 & 4/5 & 3 & three-leg stool, office chair, wardrobe \\
Mechanical & 3/3/3 & 4/5 & 3 & gear train, robot arm, bench vise \\
Vehicles & 3/3/3 & 4/5 & 3 & city bicycle, quadcopter drone, forklift \\
Tools & 3/3/3 & 4/5 & 3 & cordless drill, wristwatch, claw hammer \\
Characters & 3/3/3 & 4/5 & 3 & garden gnome, rubber duck, dragon figurine \\
Architecture & 3/3/3 & 4/5 & 3 & curbside mailbox, footbridge, garden gazebo \\
\midrule
\textbf{Total} & \textbf{18/18/18} & \textbf{24/30} & \textbf{18} & \textbf{54 items} \\
\bottomrule
\end{tabular}
\end{table}

\subsection{Baselines}
Under an input-parity rule, meaning a baseline never receives more input than
Nova3D had on an item, we compare against four families
(Table~\ref{tab:baselines}). Item sets differ by native modality: text-only
systems run the 24 text items, image-only systems the 30 image items, and
Nova3D, the naive ablation, and Meshy run all 54. CubePart is an \emph{oracle}
baseline: it cannot name segments itself, so it is handed the spec part names,
and it operates by segmenting the TripoSG mesh rather than generating geometry
of its own. Every asset, whatever its native format, is converted to GLB and
rendered through one shared headless scene (identical camera, lighting, scale
normalization).

\begin{table}[ht!]
\centering
\caption{\textbf{Baselines.} Four families under input parity. Not all rows share
an item set. CubePart is oracle-vocabulary and segments TripoSG meshes.}
\label{tab:baselines}
\begin{tabular}{lllcl}
\toprule
System & Family & Input & Items & Output representation \\
\midrule
Meshy 5~\cite{meshy2024} & mesh-native (commercial) & text+image & 54 & fused textured mesh \\
TRELLIS.2~\cite{xiang2024trellis} & mesh-native & image & 30 & fused mesh \\
TripoSG~\cite{li2025triposg} & mesh-native & image & 30 & fused mesh \\
LLaMA-Mesh~\cite{wang2024llamamesh} & mesh-native (tokens) & text & 24 & fused mesh \\
PartCrafter~\cite{lin2025partcrafter} & born-seg diffusion & image & 30 & unnamed part meshes \\
CubePart (oracle) & post-hoc seg.\ of TripoSG & image & 30 & flat labeled parts \\
MeshAnything~\cite{chen2024meshanything} & retopology & image & 30 & artist mesh \\
BlenderLLM~\cite{du2024blenderllm} & code-native & text & 24 & Blender script \\
naive same-LLM & code-native (ablation) & text+image & 54 & Blender script \\
Text2CAD~\cite{khan2024text2cad} & CAD & text & 24 & parametric STEP \\
Text-to-CadQuery~\cite{texttocadquery2025} & CAD code & text & 24 & CadQuery / STL \\
CAD-Coder~\cite{cadcoder2025} & CAD code & image & 30 & CadQuery / STEP \\
\bottomrule
\end{tabular}
\end{table}

\subsection{Evaluation axes and coverage}
The evaluation is not one universal leaderboard. Each axis uses the baselines whose
representation exposes the relevant affordance, and Table~\ref{tab:coverage} states
the item set, metric source, and explicit non-claim for each axis.

\begin{table}[ht!]
\centering
\footnotesize
\caption{\textbf{Evaluation coverage matrix.} Which baselines each claim uses, the
metric source, and what the axis does \emph{not} prove.}
\label{tab:coverage}
\begin{tabular}{L{1.5cm} L{2.5cm} L{2.0cm} L{2.4cm} L{2.4cm}}
\toprule
Axis & Methods included & Item set & Metric source & Does \emph{not} prove \\
\midrule
Reliability & Nova3D, naive same-LLM, BlenderLLM, Text2CAD, Text-to-CadQuery, CAD-Coder & modality-specific & execution + artifact logs & visual quality \\
Visual shape & Nova3D vs Meshy, TRELLIS.2, TripoSG, PartCrafter & 54/30, mixed by modality & VLM pairwise normal renders & texture SOTA \\
Textured fidelity & Nova3D vs mesh-native + code-native & modality-specific & VLM pairwise textured renders & editability / semantics \\
Naming & 13 systems & produced assets & scene-graph names + semantic judges & point-mask IoU \\
Hierarchy & 13 systems & produced assets & glTF scene graph & provable nesting correctness \\
Constraints & code-native + CAD slice & constrained items & deterministic GLB measurement & beauty / texture \\
Editability & Nova3D only & 18 local edits & deterministic gates + blinded human review & general editing SOTA \\
Articulation & all baselines (none expose native joints) & 12 articulated + all-54 readiness & joint sidecars + scene-graph actuation & perfect operating limits \\
Production readiness & Nova3D/code vs baked GLB; part-aware where applicable & varies by metric & UV, primitive residual, material telemetry & artist-final quality \\
\bottomrule
\end{tabular}
\end{table}

\section{Code-Native Reliability and the System Package}
\label{sec:reliability}
\begin{figure}[ht!]
\centering
\begin{subfigure}{0.5\linewidth}
\includegraphics[width=\linewidth]{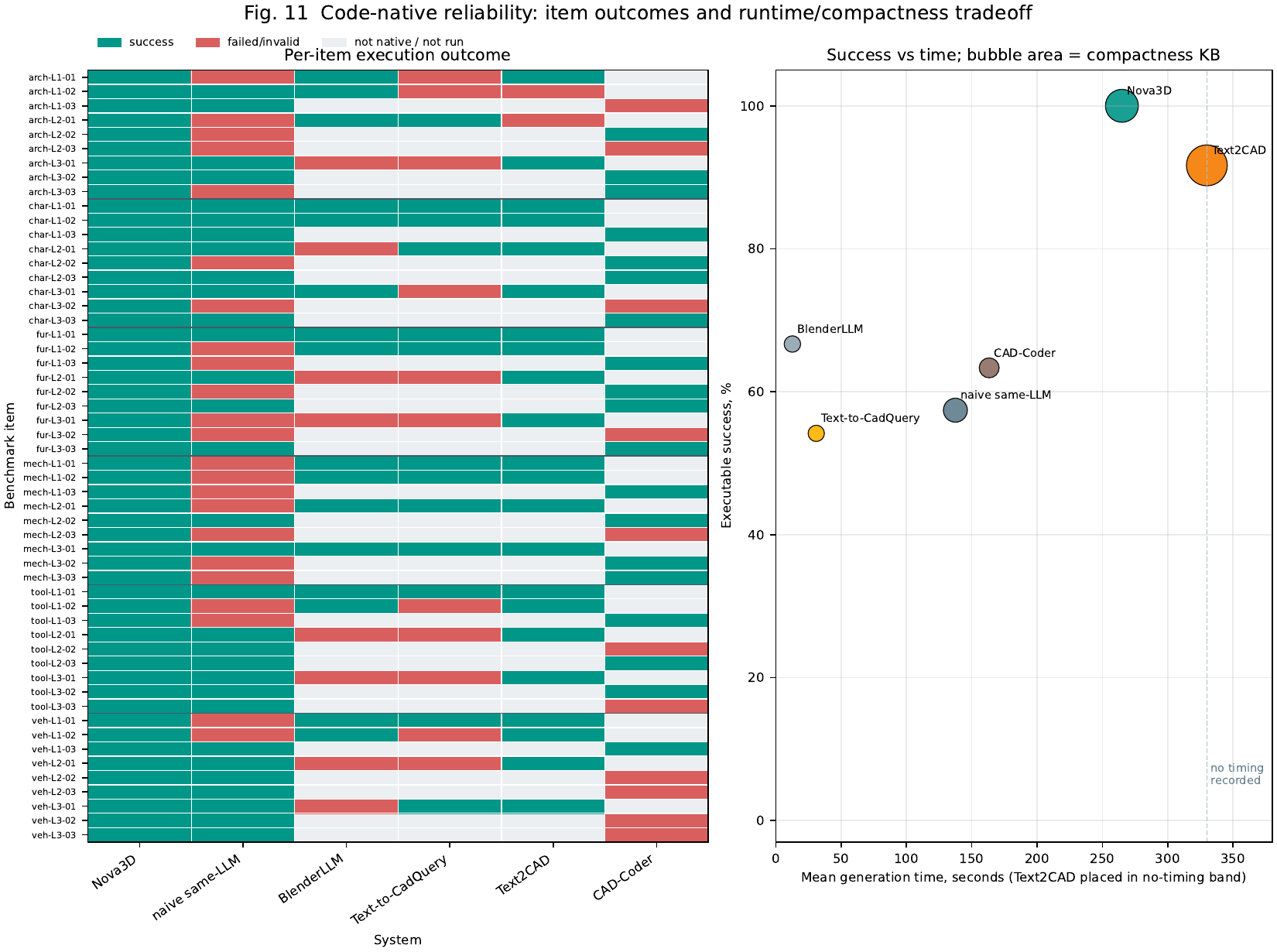}
\caption{Reliability matrix}
\end{subfigure}\hfill
\begin{subfigure}{0.48\linewidth}
\includegraphics[width=\linewidth]{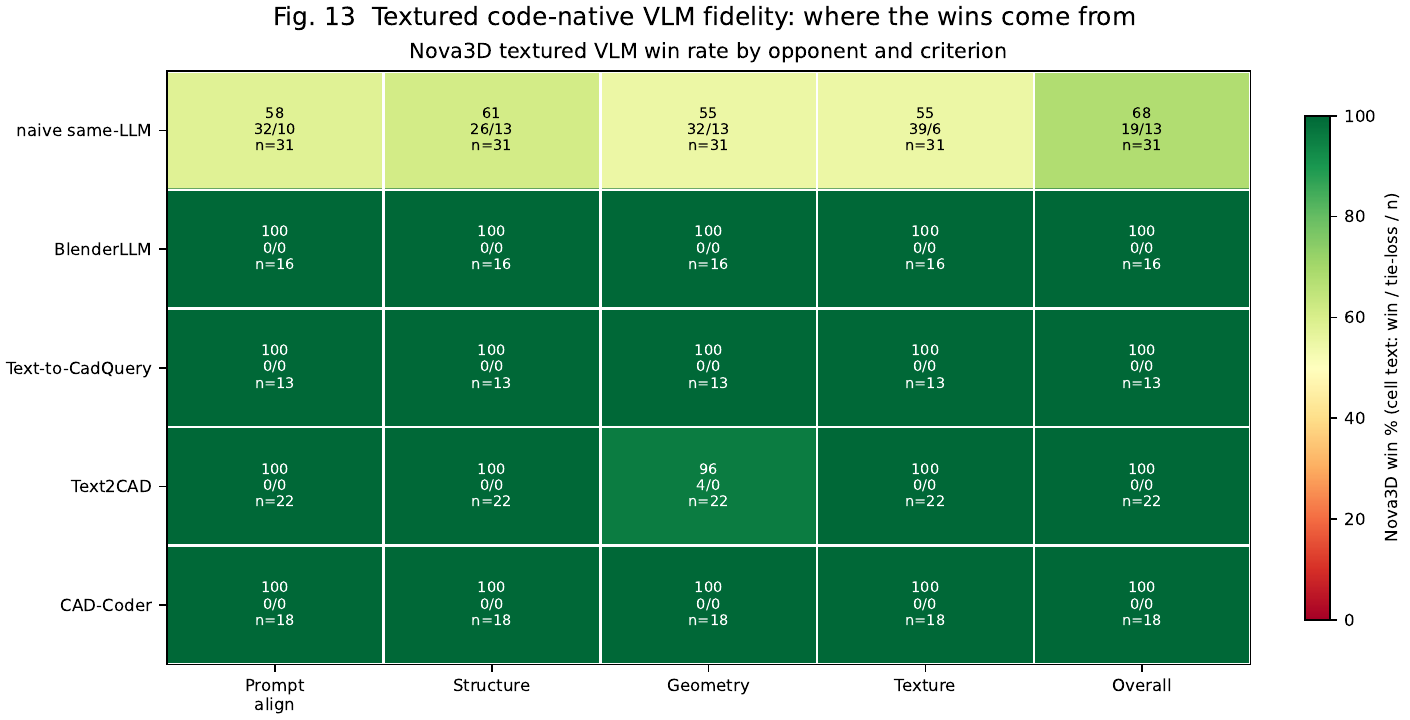}
\caption{Textured fidelity vs code-native}
\end{subfigure}
\caption{\textbf{The system package, not the base model.} (a) Per-item execution
outcome plus success/time/compactness; only Nova3D runs and saves an artifact for
all 54 items, while the same base model without the package fails or invalidates
nearly half. (b) Nova3D win-rate against each code-native opponent by criterion
(textured, both-orders-resolved); its closest competitor is the naive ablation
(overall 67.7/19.4/12.9 win/tie/loss).}
\label{fig:reliability}
\end{figure}

The first link in the chain is reliability, and it answers the sharpest objection
up front: \emph{is this just prompting an LLM to write Blender?} It is not. Nova3D
is the only evaluated code-native system that produces an executable program and a
valid artifact for all 54 mixed-modality items: \textbf{100\% executable, 100\%
artifact-saved}. The same base model without the Nova3D system package (the naive
same-LLM ablation) succeeds on only 31/54; the published BlenderLLM on 16/24; and
the CAD baselines range from 13/24 to 22/24 (Table~\ref{tab:reliability}, Figure~\ref{fig:reliability}). Item
sets differ by modality and are reported, not assumed identical. Crucially,
Text2CAD's high basic validity is reconstruction success, not visual or semantic
success: it typically emits a crude fused solid that ignores most of the prompt.

\begin{table}[ht!]
\centering
\caption{\textbf{Code-native reliability and compactness.} Executable and
artifact-saved rates on each system's native item set.}
\label{tab:reliability}
\begin{tabular}{llrrrl}
\toprule
System & Item set & N & Executable & Artifact saved & Compactness \\
\midrule
\textbf{Nova3D} & text+image (54) & 54 & \textbf{100.0\%} & \textbf{100.0\%} & 489 LOC / 18.7\,KB \\
naive same-LLM & text+image (54) & 54 & 57.4\% & 57.4\% & 210 LOC / 7.9\,KB \\
BlenderLLM & text (24) & 24 & 66.7\% & 66.7\% & 41 LOC / 1.4\,KB \\
Text-to-CadQuery & text (24) & 24 & 54.2\% & 54.2\% & 51 LOC / 1.3\,KB \\
Text2CAD & text (24) & 24 & 91.7\% & 91.7\% & STEP avg 31.3\,KB \\
CAD-Coder & image (30) & 30 & 63.3\% & 60.0\% & 52 LOC / 4.1\,KB \\
\bottomrule
\end{tabular}
\end{table}

\paragraph{Systems telemetry (compact).}
Under a single fixed configuration, Nova3D reaches a 100\% final success rate
(all 54 items repaired within $\le3$ iterations) with an \textbf{81.5\%
first-shot rate} (44 items needed no repair, 8 needed one, 2 needed two). Median
wall time is 218.5\,s and median token usage 74.5k; assets carry a median of 31
mesh-bearing parts (up to 339). First-shot rates hold across domains and
difficulty (L1 100\%, L2 67\%, L3 78\%), indicating that the harness, not
per-domain engineering, drives reliability. Full telemetry is in the appendix.

\section{Visual and Geometry Quality}
\begin{figure}[ht!]
\centering
\setlength{\tabcolsep}{2pt}
\renewcommand{\arraystretch}{0.6}
\begin{tabular}{L{1.7cm}cc}
& \small\textbf{Nova3D (ours)} & \small\textbf{Mesh-native} \\
\small Architecture (structured win) &
\includegraphics[width=0.36\linewidth]{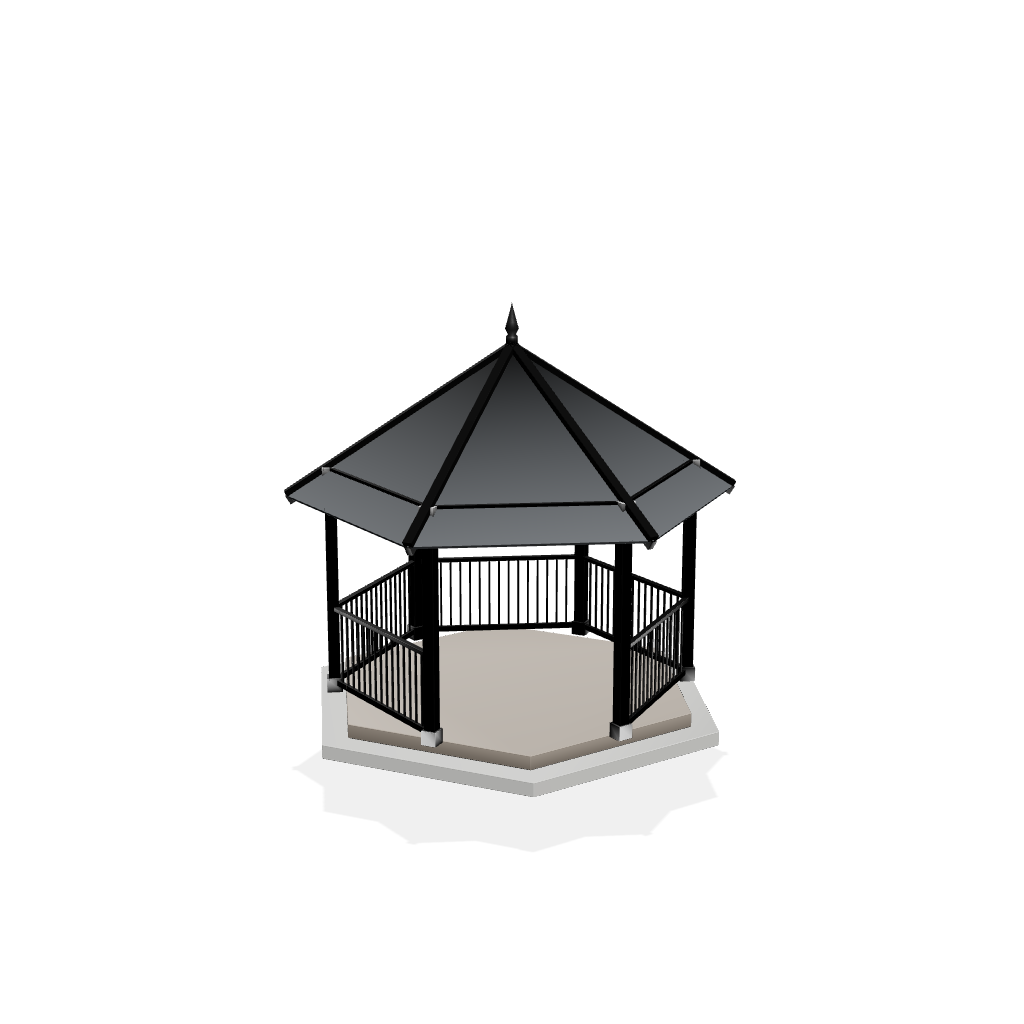} &
\includegraphics[width=0.36\linewidth]{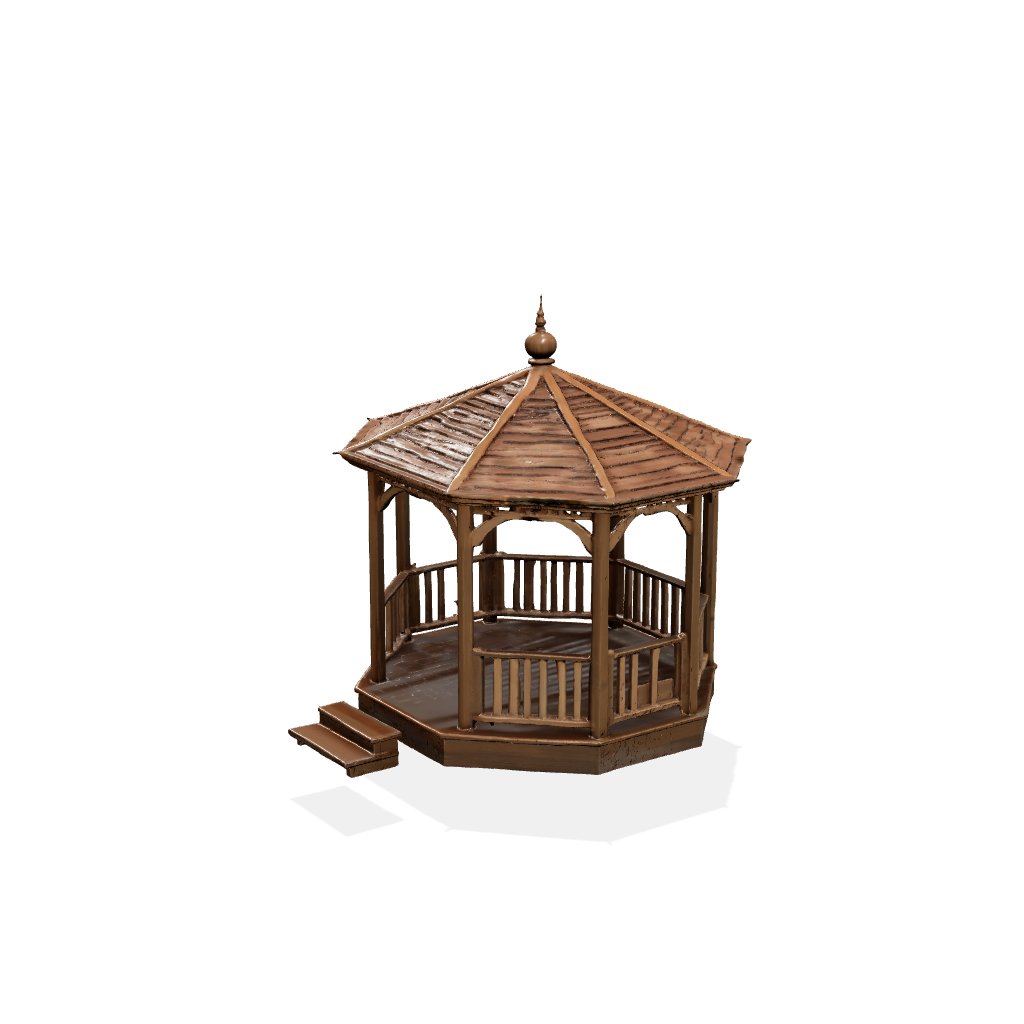} \\
\small Vehicles (win) &
\includegraphics[width=0.36\linewidth]{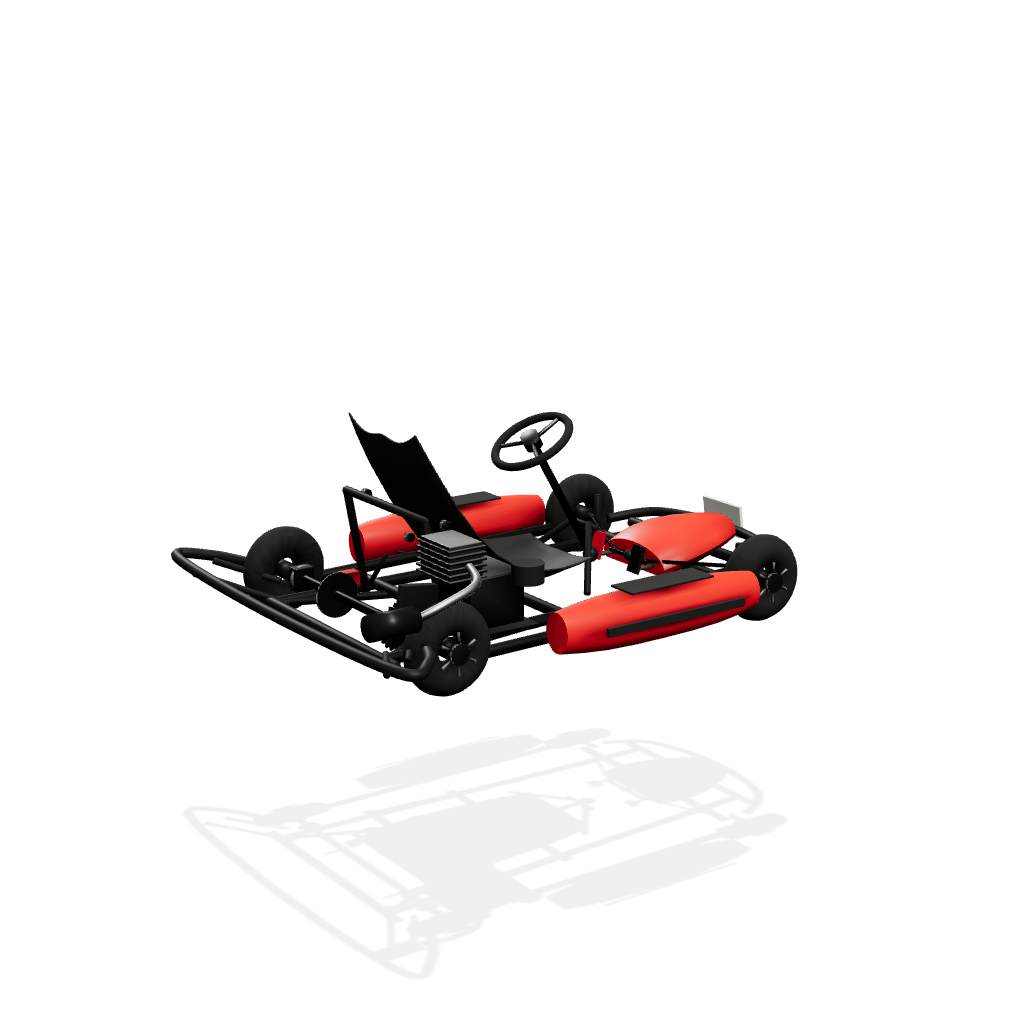} &
\includegraphics[width=0.36\linewidth]{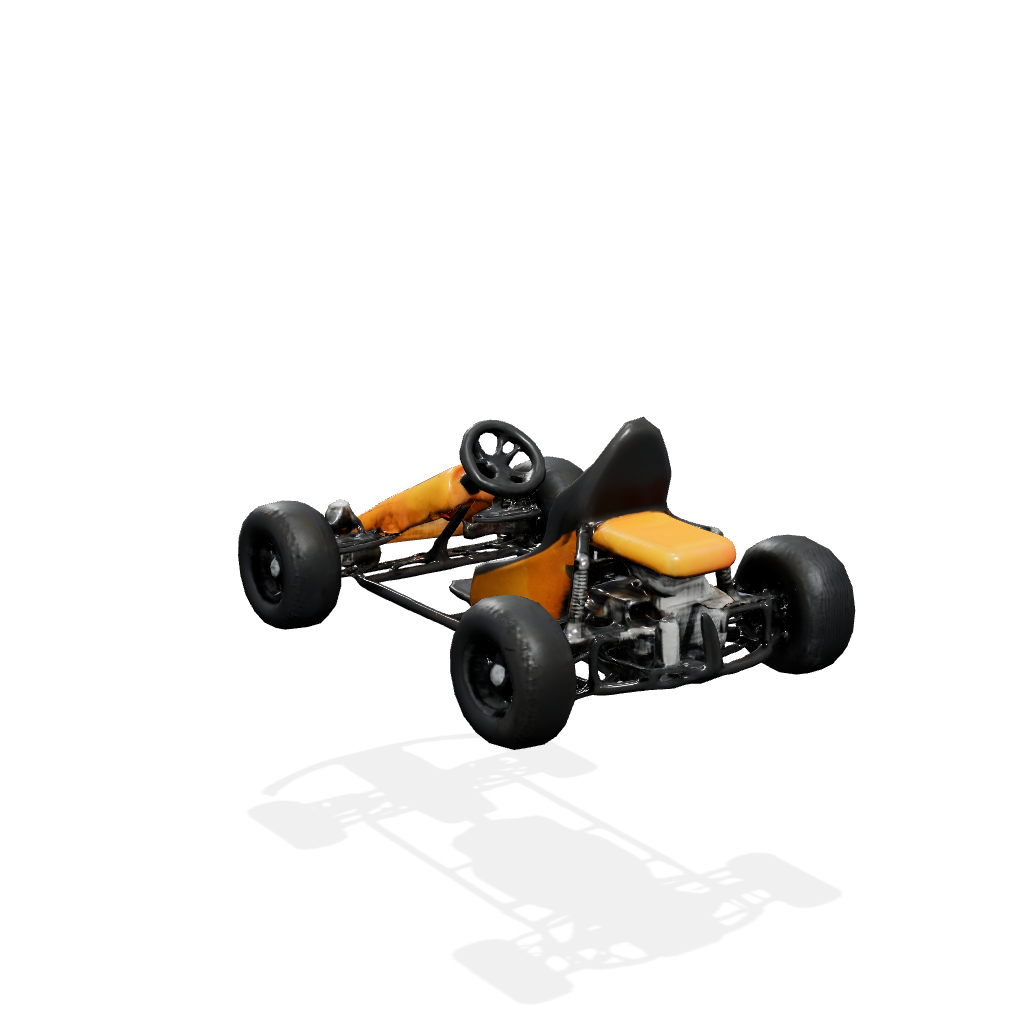} \\
\small Characters (conceded) &
\includegraphics[width=0.36\linewidth]{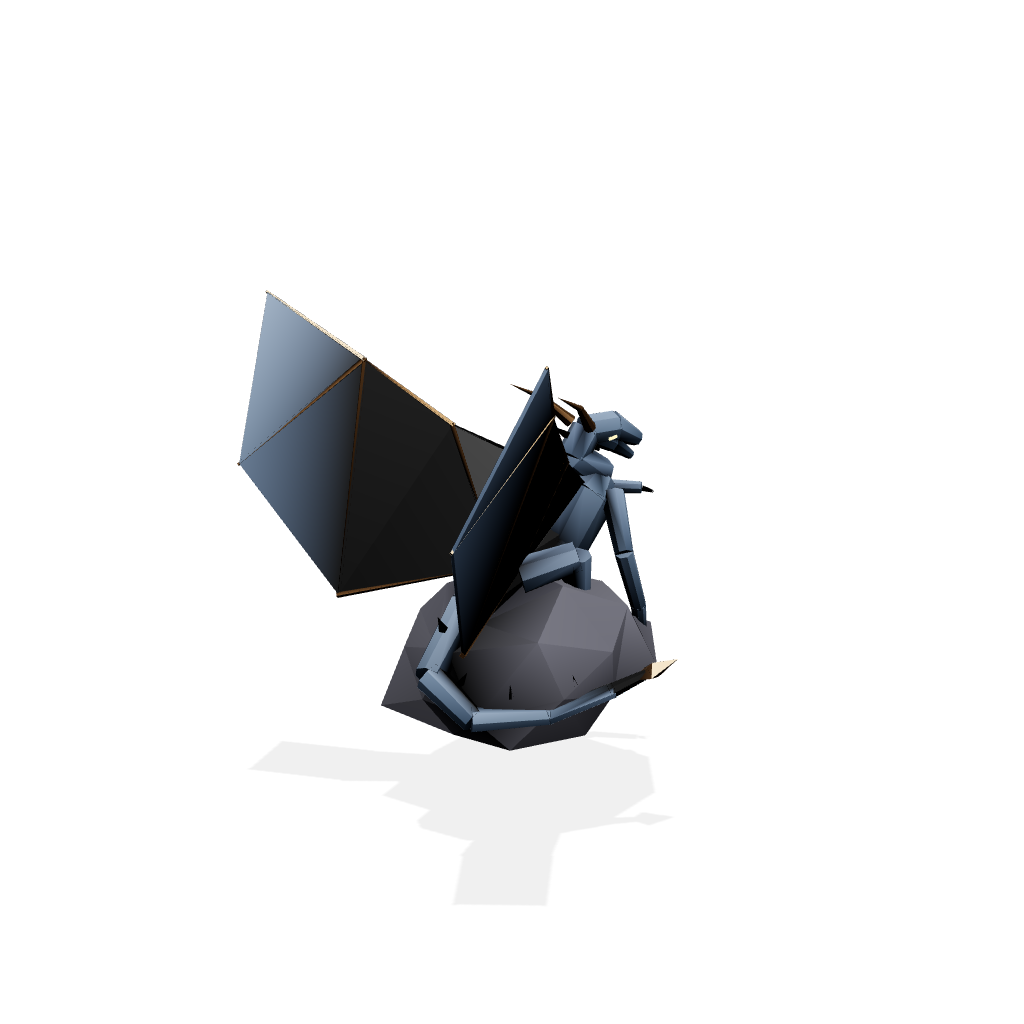} &
\includegraphics[width=0.36\linewidth]{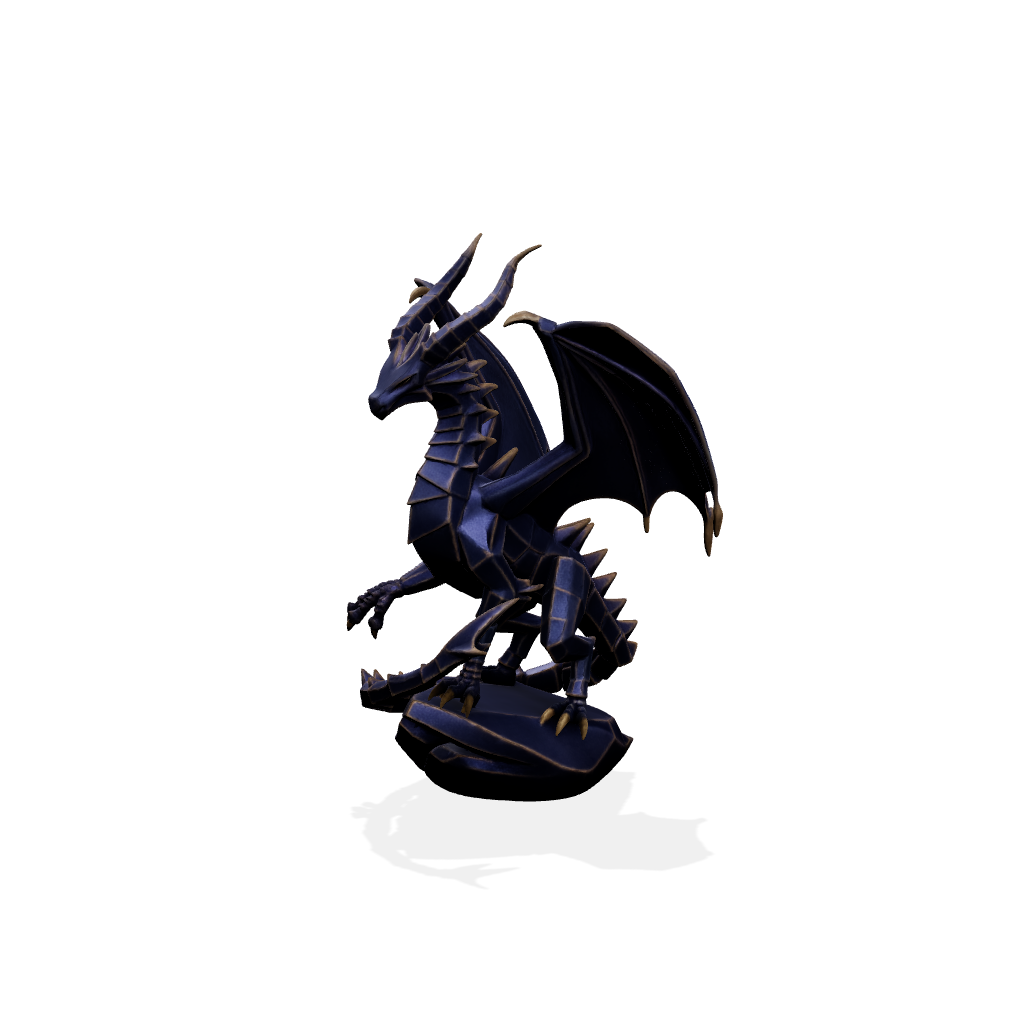} \\
\end{tabular}
\caption{\textbf{Honest qualitative comparison} (as-delivered renders, shared
scene). Nova3D holds the structured domains (top: a garden gazebo with clean
panels and posts) and wins vehicles (middle: a go-kart, 55.6\% domain win-rate),
while it concedes freeform organic surface detail and baked-texture realism to
the strongest mesh-native model on characters (bottom: a dragon). Right column
is Meshy (top, middle) and TRELLIS.2 (bottom).}
\label{fig:qualitative}
\end{figure}

The second link answers a different reviewer question, namely whether the
structure costs asset quality, and it is deliberately a defensive check, not the
headline. The answer is that Nova3D's assets are competitive on shape, conceding
texture realism to baked-PBR systems (Figure~\ref{fig:qualitative}).

\subsection{Perceptual geometry}
\begin{figure}[ht!]
\centering
\begin{subfigure}{0.33\linewidth}
\includegraphics[width=\linewidth]{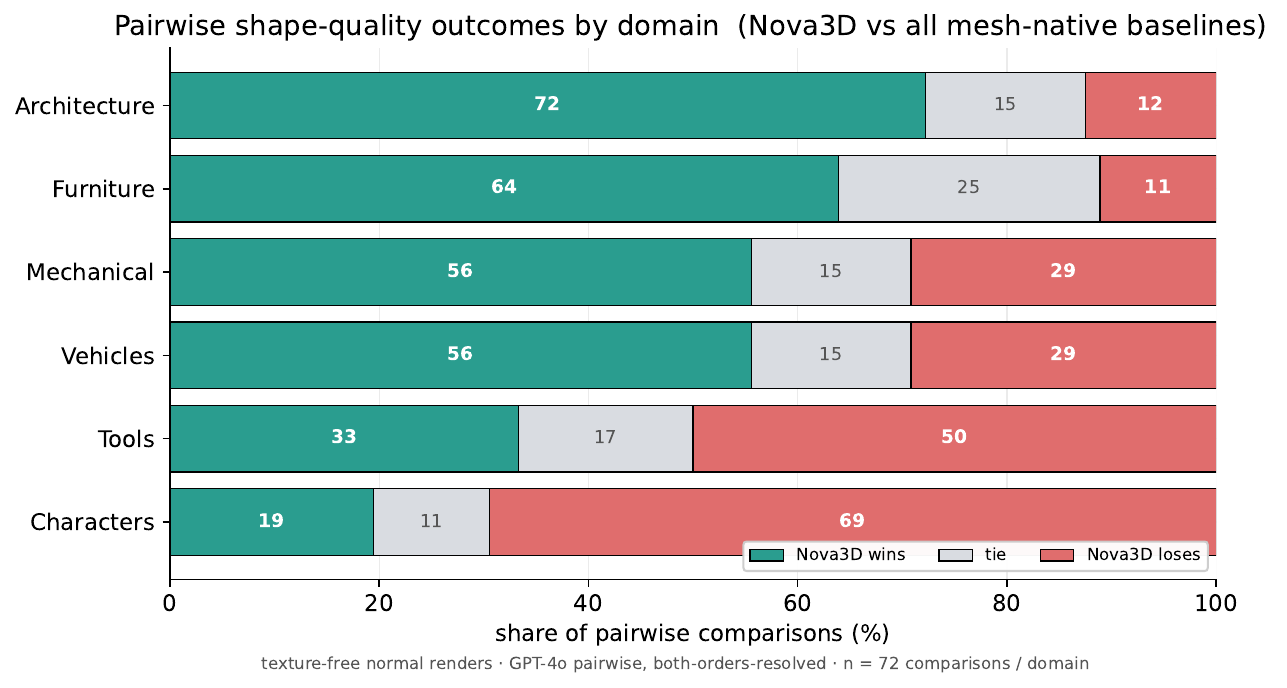}
\caption{By domain}
\end{subfigure}\hfill
\begin{subfigure}{0.33\linewidth}
\includegraphics[width=\linewidth]{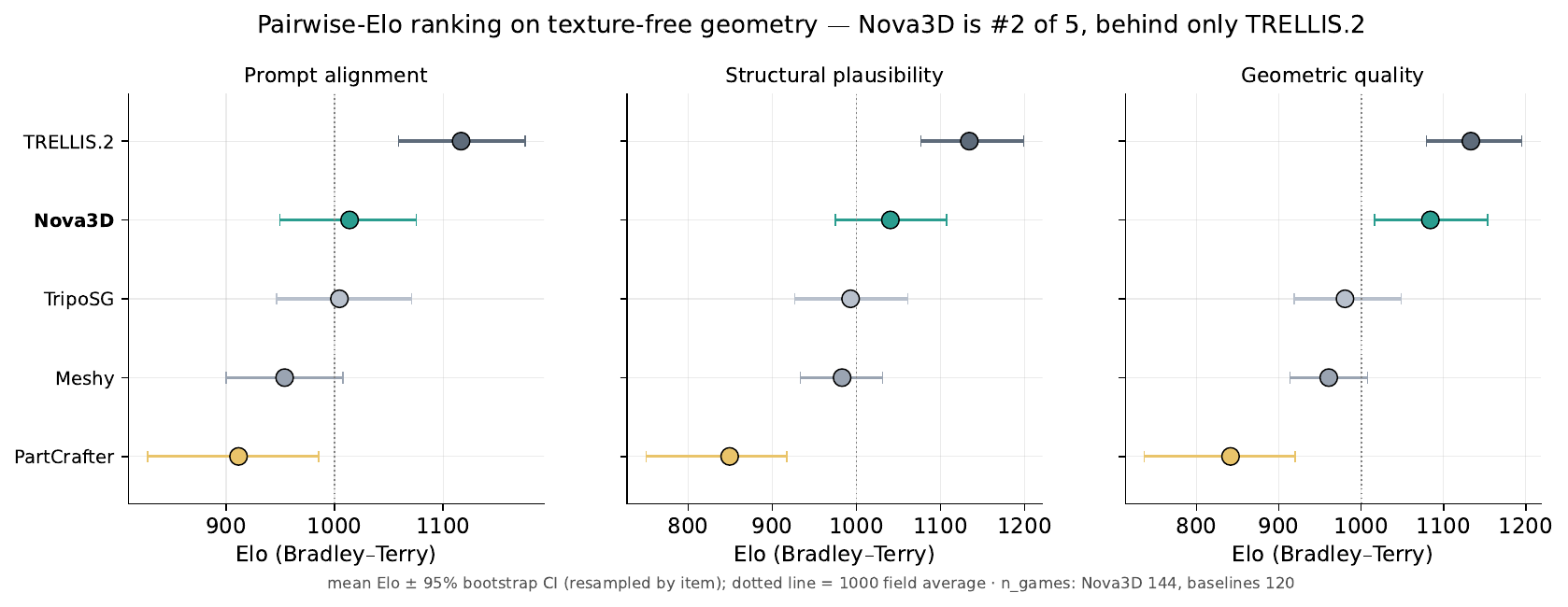}
\caption{Elo ranking}
\end{subfigure}\hfill
\begin{subfigure}{0.31\linewidth}
\includegraphics[width=\linewidth]{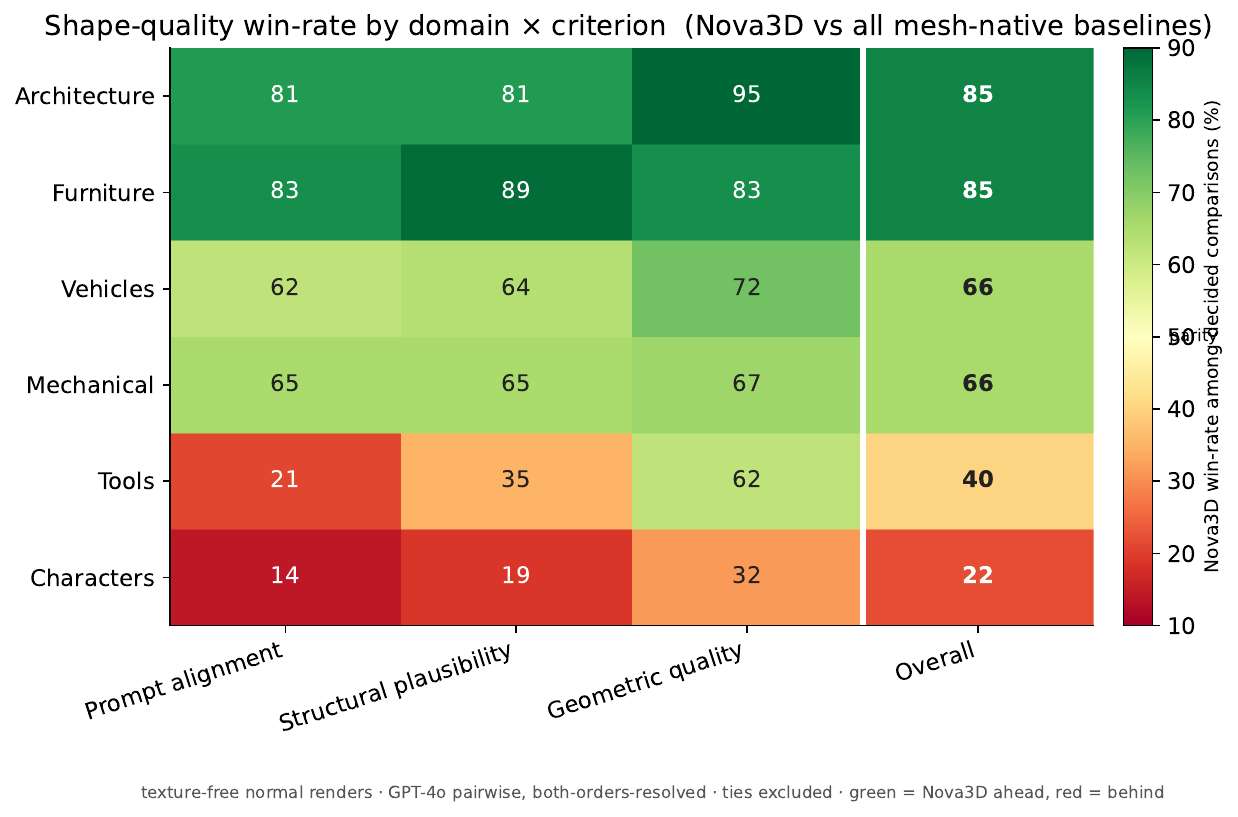}
\caption{Domain $\times$ criterion}
\end{subfigure}
\caption{\textbf{Perceptual fidelity on texture-free geometry.} Nova3D wins the
structured domains and is \#2 of 5 in Elo, behind only TRELLIS.2. Panel (c) shows
that geometric quality is Nova3D's strongest criterion in every domain, including
the domains it loses (62\% in tools); the losses concentrate in prompt alignment
and structural plausibility, i.e.\ in matching the object's identity and
completeness, not in surface quality.}
\label{fig:fidelity}
\end{figure}

We run a GPTEval3D-style pairwise tournament: every asset, every system, is
rendered through one identical headless scene as a texture-free surface-normal
map (front and back, 1024\,px, tight-cropped) to isolate shape, and GPT-4o
compares two anonymized assets on prompt alignment, structural plausibility, and
geometric quality. Each pair is judged in both A/B orders and a win is counted
only when both agree; all 324 pairs (648 judgments) completed. Aggregated,
Nova3D wins geometric quality outright (55.6\% win / 27.1\% loss) and is
even-to-ahead on the other two criteria. The result is strongly
domain-dependent: Nova3D wins all four structured domains and loses tools and
characters (Table~\ref{tab:fidelity}, Figure~\ref{fig:fidelity}). The losses are not a surface-quality
failure. Geometric quality is Nova3D's strongest criterion in \emph{every}
domain, including the two it loses (62\% in tools, 32\% in characters); the
deficits concentrate in prompt alignment and structural plausibility, that is,
in reproducing the object's identity and completeness rather than in surface
quality. By opponent, Nova3D is even with commercial Meshy and ahead of TripoSG
and PartCrafter, and it trails only TRELLIS.2, placing second of five in
Bradley--Terry Elo on every criterion. We do not claim state-of-the-art
fidelity; the point is parity-or-better shape quality alongside structure the
other methods do not natively expose.

\begin{table}[ht!]
\centering
\caption{\textbf{Pairwise shape-quality by domain} (Nova3D POV, averaged over three
criteria, both-orders-resolved, $n=72$/domain; texture-free normal renders, GPT-4o).}
\label{tab:fidelity}
\begin{tabular}{lrrrl}
\toprule
Domain & Win & Tie & Loss & Verdict \\
\midrule
Architecture & 72.2\% & 15.3\% & 12.5\% & strong win \\
Furniture & 63.9\% & 25.0\% & 11.1\% & strong win \\
Mechanical & 55.6\% & 15.3\% & 29.2\% & win \\
Vehicles & 55.6\% & 15.3\% & 29.2\% & win \\
Tools & 33.3\% & 16.7\% & 50.0\% & loss (alignment-driven) \\
Characters & 19.4\% & 11.1\% & 69.4\% & loss (alignment-driven) \\
\bottomrule
\end{tabular}
\end{table}

\subsection{Textured (as-delivered) disclosure}
Because delivered assets include appearance, we additionally judge as-delivered
renders under neutral image-based lighting, adding texture-quality and overall
criteria. The geometry verdict is stable with textures visible (geometric
quality 56.2\%, essentially identical to the normals track), confirming it is
not a render artifact. As expected, Nova3D concedes texture realism: per
opponent (which we report instead of the PartCrafter-inflated aggregate), it
loses texture quality to Meshy and TRELLIS.2 and is even with TripoSG
(Table~\ref{tab:textured}). The honest statement is that code-native concedes
texture realism to baked-PBR mesh-native methods while keeping its geometry
edge.

\begin{table}[ht!]
\centering
\caption{\textbf{Textured/as-delivered fidelity, per opponent} (Nova3D win-rate,
both-orders-resolved). Reported per opponent because the aggregate is inflated by
PartCrafter.}
\label{tab:textured}
\begin{tabular}{lrrr}
\toprule
Opponent & Geometric quality & Texture quality & Overall \\
\midrule
Meshy (commercial) & 54\% & 19\% & 39\% \\
TRELLIS.2 & 43\% & 23\% & 37\% \\
TripoSG & 43\% & 43\% & 43\% \\
PartCrafter & 87\% & 90\% & 83\% \\
\bottomrule
\end{tabular}
\end{table}

\subsection{Geometry integrity}
\begin{figure}[ht!]
\centering
\begin{subfigure}{0.48\linewidth}
\includegraphics[width=\linewidth]{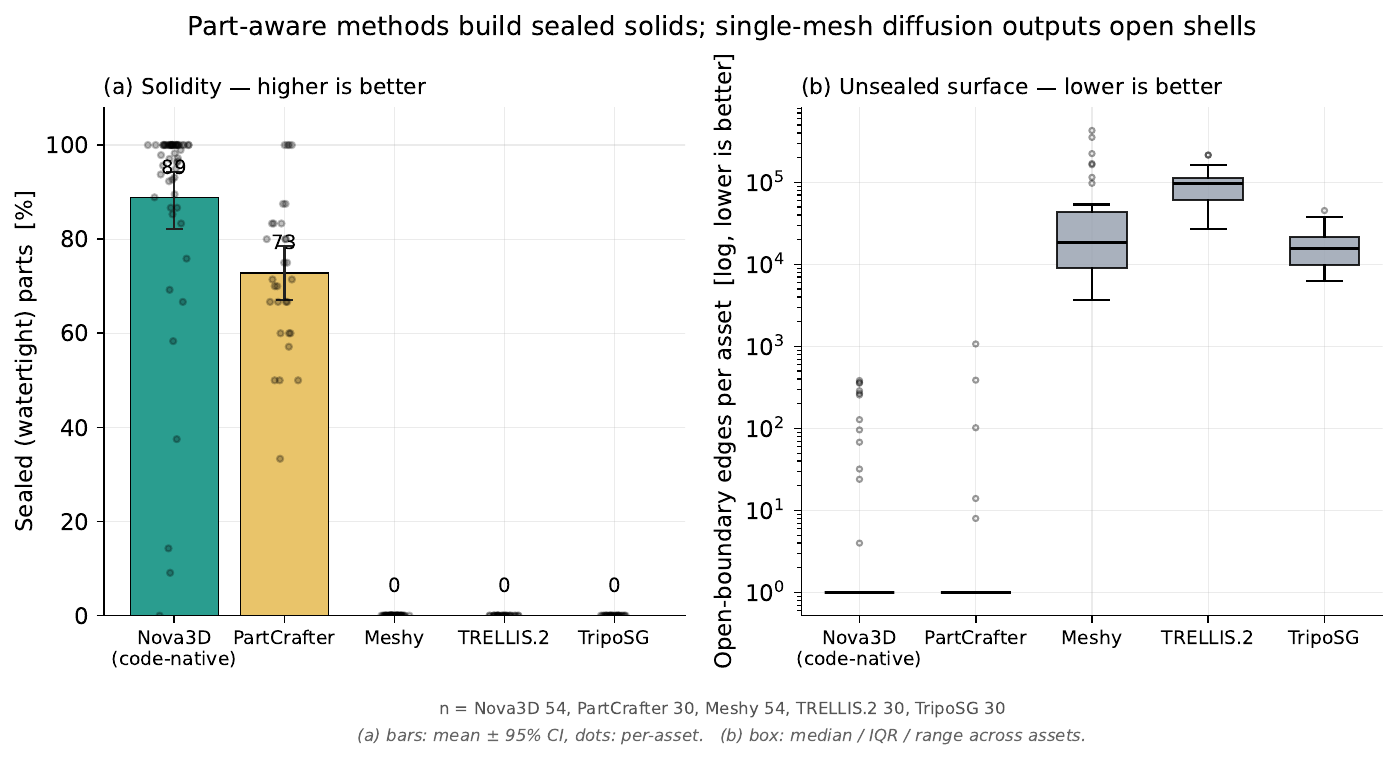}
\caption{Integrity}
\end{subfigure}\hfill
\begin{subfigure}{0.48\linewidth}
\includegraphics[width=\linewidth]{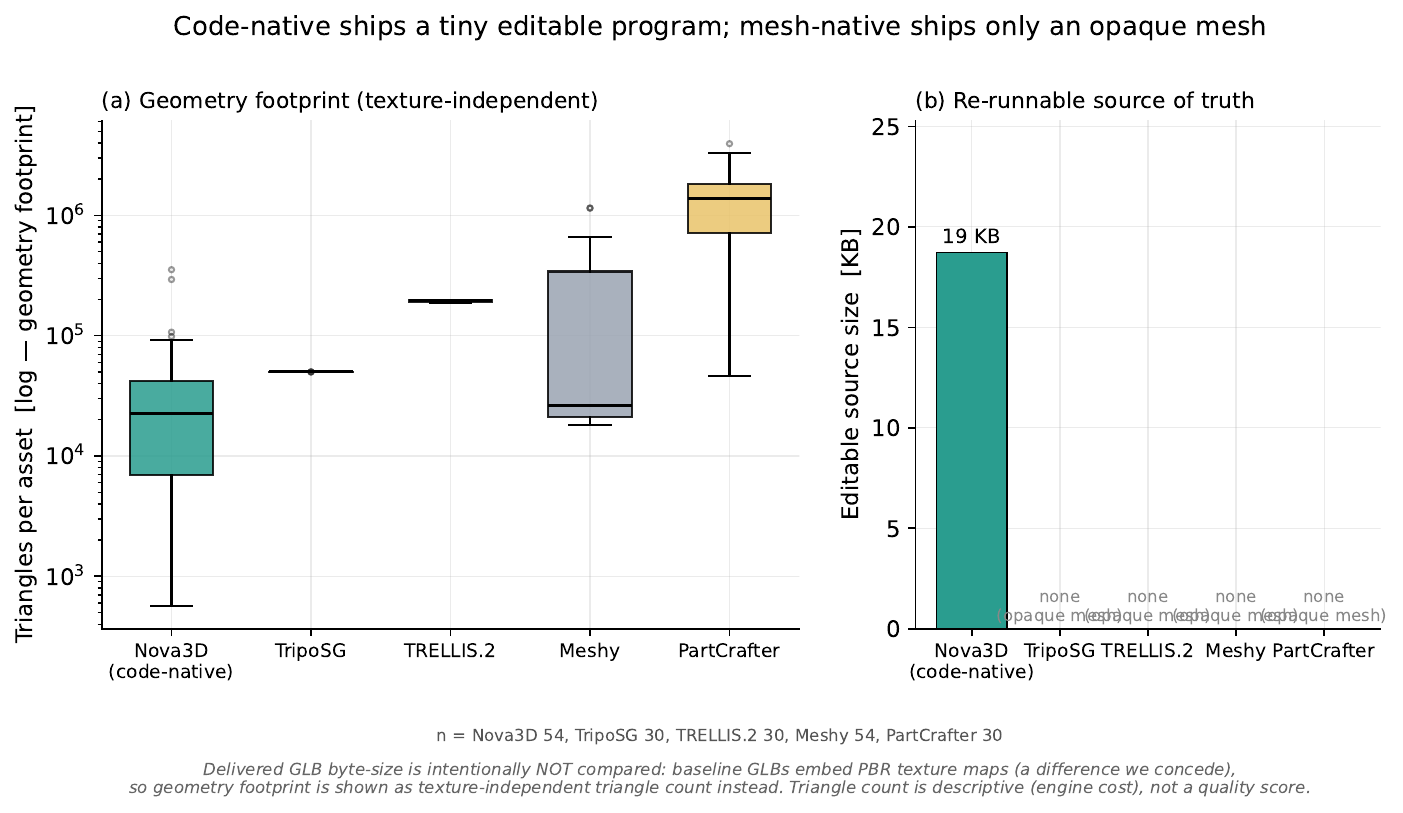}
\caption{Compactness}
\end{subfigure}
\caption{\textbf{Part-aware methods build sealed solids; single-mesh diffusion
outputs open shells; only code-native ships a tiny editable program.} Delivered
GLB byte-size is not compared as a geometry win because baseline GLBs embed PBR
texture maps; compactness is argued on triangle count and editable-source size.}
\label{fig:geometry}
\end{figure}

Reference-free integrity metrics, computed fairly across single-mesh and
multi-part outputs (a seam-merged copy per component), show a sharp split by
representation. Single-mesh diffusion produces open shells: \textbf{0\%
watertight}, with 17{,}000--97{,}000 open-boundary edges per asset. Part-structured
methods produce sealed solids, and among them Nova3D is cleanest at
\textbf{88.8\% watertight and 91.0\% manifold} (median $\sim0$ open edges);
PartCrafter reaches 72.8\%/76.0\% (Table~\ref{tab:geometry}, Figure~\ref{fig:geometry}). Nova3D also
produces the lightest geometry ($\sim$39k triangles, 1.16\,MB) and, uniquely, an
editable source of only \textbf{18.7\,KB}, the description-length argument made
concrete. The clean split is single-mesh vs part-structured, so we claim
``cleanest among part-aware methods,'' not unique solidity; and the single-mesh
failure is sealing, not manifoldness (TripoSG is 0.90 manifold yet 0.00
watertight).

\begin{table}[ht!]
\centering
\caption{\textbf{Geometry integrity and compactness (reference-free).} N:
Nova3D/Meshy 54, others 30. We do not report triangle-shape/dihedral metrics: they
rank by tessellation density, not quality, and would invert the conclusion (a 1.39M-triangle
output tops them).}
\label{tab:geometry}
\begin{tabular}{lrrrrrr}
\toprule
Method & Watertight $\uparrow$ & Manifold $\uparrow$ & Open edges $\downarrow$ & Triangles & GLB size $\downarrow$ & Source $\downarrow$ \\
\midrule
\textbf{Nova3D} & \textbf{0.89} & \textbf{0.91} & \textbf{42} & \textbf{39k} & 1.16\,MB & \textbf{18.7\,KB} \\
PartCrafter & 0.73 & 0.76 & 53 & 1{,}394k & 27.9\,MB & --- \\
TripoSG & 0.00 & 0.90 & 17{,}664 & 50k & 2.8\,MB & --- \\
Meshy & 0.00 & 0.67 & 47{,}319 & 213k & 12.5\,MB & --- \\
TRELLIS.2 & 0.00 & 0.37 & 96{,}840 & 194k & 16.4\,MB & --- \\
\bottomrule
\end{tabular}
\end{table}

\section{Native Semantic Structure}
\label{sec:structure}
This is the core of the paper: the code representation exposes named parts and a
real assembly tree at generation time. These are two distinct claims, whether
parts are \emph{named} and whether they are organized into a \emph{tree}, and we
evaluate them separately.

\subsection{Semantic naming}
\begin{figure}[ht!]
\centering
\begin{subfigure}{0.46\linewidth}
\includegraphics[width=\linewidth]{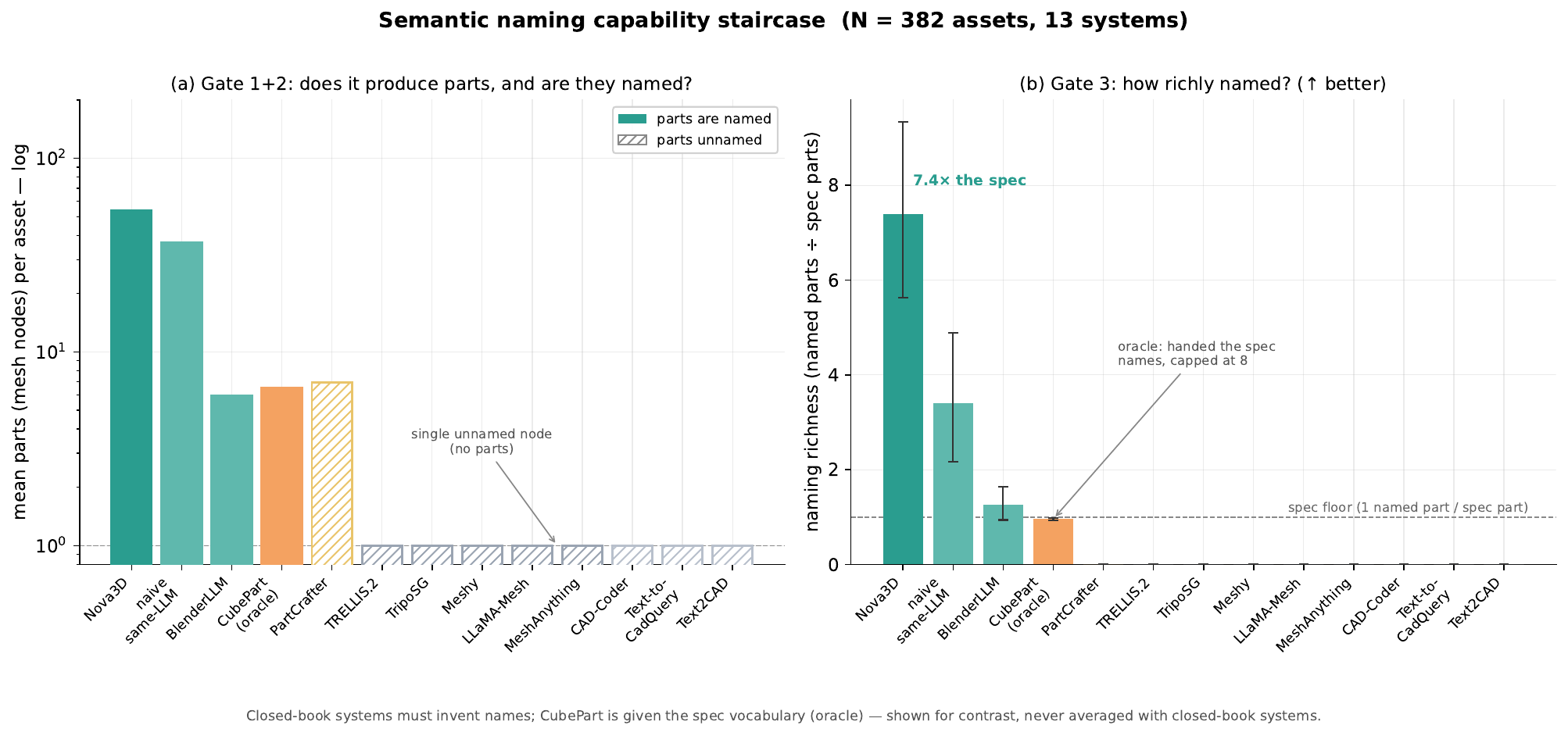}
\caption{Capability staircase}
\end{subfigure}\hfill
\begin{subfigure}{0.46\linewidth}
\includegraphics[width=\linewidth]{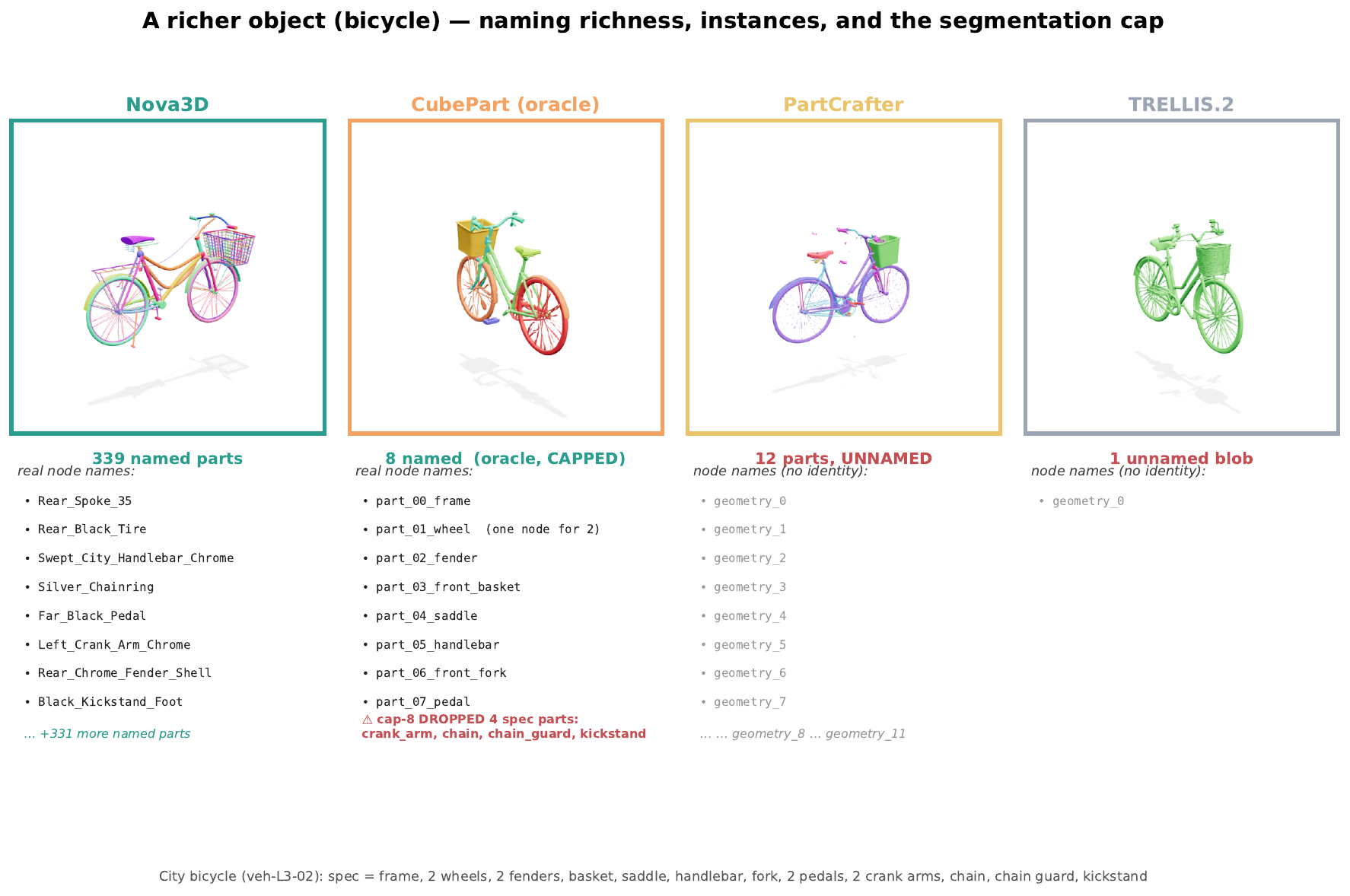}
\caption{Worked example (city bicycle)}
\end{subfigure}
\caption{\textbf{Native semantic handles.} (a) The parts $\to$ named $\to$ richness
staircase across 13 systems (bootstrap CIs). (b) A city bicycle: Nova3D exposes
hundreds of individually named parts (spokes, crank arms); oracle CubePart is
capped at 8 and drops four real spec parts; the part-aware diffusion model returns
\texttt{geometry\_N}; the mesh-native model returns a single blob.}
\label{fig:naming}
\end{figure}

We order systems as a capability staircase (produces parts $\to$ named $\to$
\emph{richness}) and lead with richness, meaning meaningful named parts divided
by spec parts, because it counts structure and cannot be gamed by copying the
spec's words. Nova3D exposes \textbf{7.4$\times$} the spec's parts with full
nameability, versus $3.4\times$ for the naive ablation, $1.26\times$ for
BlenderLLM, $0.96\times$ for the oracle CubePart (capped), and 0 for mesh-native
and CAD baselines (Table~\ref{tab:naming}, Figure~\ref{fig:naming}). It also individually names repeated
instances (separate teeth, spokes, legs), reflected in its
instance-discriminability score. CubePart is marked oracle in every table and is
never pooled with closed-book systems: it was handed the spec vocabulary, so its
high recall is expected.

\begin{table}[ht!]
\centering
\caption{\textbf{Semantic naming capability staircase} (means over produced assets).
Richness is the headline; recall is a parroting-biased lower bound (\dag{} CubePart
was handed the spec strings, so its high recall is expected).}
\label{tab:naming}
\begin{tabular}{llrrrrr}
\toprule
Method & Condition & Parts (\#) & Named? & \textbf{Richness} & Instance & Recall\dag \\
\midrule
\textbf{Nova3D} & closed-book & 53.5 & 1.00 & \textbf{7.40} & \textbf{0.46} & 0.81 \\
naive same-LLM & closed-book & 36.5 & 0.84 & 3.40 & 0.25 & 0.60 \\
BlenderLLM & closed-book & 5.9 & 0.98 & 1.26 & 0.36 & 0.71 \\
CubePart & \emph{oracle} & 6.5 & 1.00 & 0.96 & 0.01 & 0.96\dag \\
PartCrafter & closed-book & 7.0 & 0.00 & 0.00 & 0.00 & 0.00 \\
Mesh-native / CAD & --- & 1.0 & --- & 0.00 & 0.00 & 0.00 \\
\bottomrule
\end{tabular}
\end{table}

\subsubsection*{Judge-validated semantic correctness}
Because generated scene graphs do not share point-level ground-truth masks, we
evaluate semantic naming with exported node names, reference part lists, and a
multi-judge semantic-matching protocol; this is \emph{not} segmentation IoU, and
it validates whether native names denote real object parts. It preempts the
natural objection that the names are just plausible-sounding hallucinations.
Under a three-judge consensus (Table~\ref{tab:namingjudge}), Nova3D's
closed-book semantic recall is \textbf{0.82} with precision \textbf{0.99}
(almost no hallucinated parts) and roughly \textbf{14} judge-confirmed real
parts beyond the spec per asset; oracle CubePart out-recalls on the spec words
but exposes no extra real parts. The judges are consistent with one another:
mean inter-judge recall spread is 0.141 (median 0.0) and 70\% of assets fall
within a 0.20 spread.

\begin{table}[ht!]
\centering
\caption{\textbf{Semantic-judge reliability and results.} A multi-judge matching
protocol was the only available way to score native names in the absence of shared
point-level masks.}
\label{tab:namingjudge}
\begin{tabular}{lr}
\toprule
Item & Value \\
\midrule
Verdicts & 387 \\
Judge models & Claude Opus 4.8, GPT-5, Gemini 3.1 Pro \\
Nova3D recall & 0.82 \\
Nova3D precision & 0.99 \\
Nova3D extra real parts & 14.0 \\
Inter-judge recall mean spread & 0.141 \\
Inter-judge recall median spread & 0.0 \\
Assets within 0.20 recall spread & 70\% \\
\bottomrule
\end{tabular}
\end{table}

\subsection{Assembly hierarchy}
Naming is not hierarchy: a system can name parts and still export them flat
(BlenderLLM names objects but emits no tree; CubePart labels regions but returns
a flat list). This axis asks whether parts are organized into a \emph{tree} of
named sub-assemblies and joint pivots, the structure an engine, animator, or
agent actually edits. We read every metric deterministically from the exported
glTF scene graph. A flat mesh has semantic depth 1; unnamed exporter wrapper
nodes are collapsed so they cannot fake hierarchy. Nova3D emits a real assembly
tree for \textbf{100\% of assets} (mean depth 3.7, $\approx$6 named groups,
$\approx$8 pivots). Every non-code-native mesh/CAD/segmentation/retopology
baseline is flat, while the naive same-LLM code ablation produces a tree only
inconsistently (45\% of assets), so the consistent, deep tree is a product of
the Nova3D system package, not of an LLM writing code per se
(Table~\ref{tab:hierarchy}). We score that a real, named, deep tree with pivots
\emph{exists}; we do not put a number on whether every nesting decision is
provably semantically correct (an LLM-judge correctness rung was designed and
dropped to keep the axis deterministic).

\begin{table}[ht!]
\centering
\caption{\textbf{Assembly-tree structure} (deterministic, from the glTF scene graph).
CubePart is oracle-vocabulary; all others closed-book. The naive ablation is
separated from the non-code baselines because it is partially hierarchical.}
\label{tab:hierarchy}
\begin{tabular}{llrrrrr}
\toprule
Method & Type & N & Has tree & Depth & Named groups & Pivots \\
\midrule
\textbf{Nova3D} & code-native & 54 & \textbf{1.00} & \textbf{3.69} & \textbf{6.20} & \textbf{8.17} \\
naive same-LLM (ablation) & code-native & 31 & 0.45 & 1.55 & 1.03 & 1.10 \\
BlenderLLM & code-native & 16 & 0.00 & 1.0 & 0 & 0 \\
CubePart & oracle seg.\ of TripoSG & 30 & 0.00 & 1.0 & 0 & 0.2 \\
PartCrafter & born-seg diffusion & 30 & 0.00 & 1.0 & 0 & 0 \\
mesh-native / CAD / retopo & various & 24--54 & 0.00 & 1.0 & 0 & 0 \\
\bottomrule
\end{tabular}
\end{table}

\section{Constraint Satisfaction}
\begin{figure}[ht!]
\centering
\includegraphics[width=0.78\linewidth]{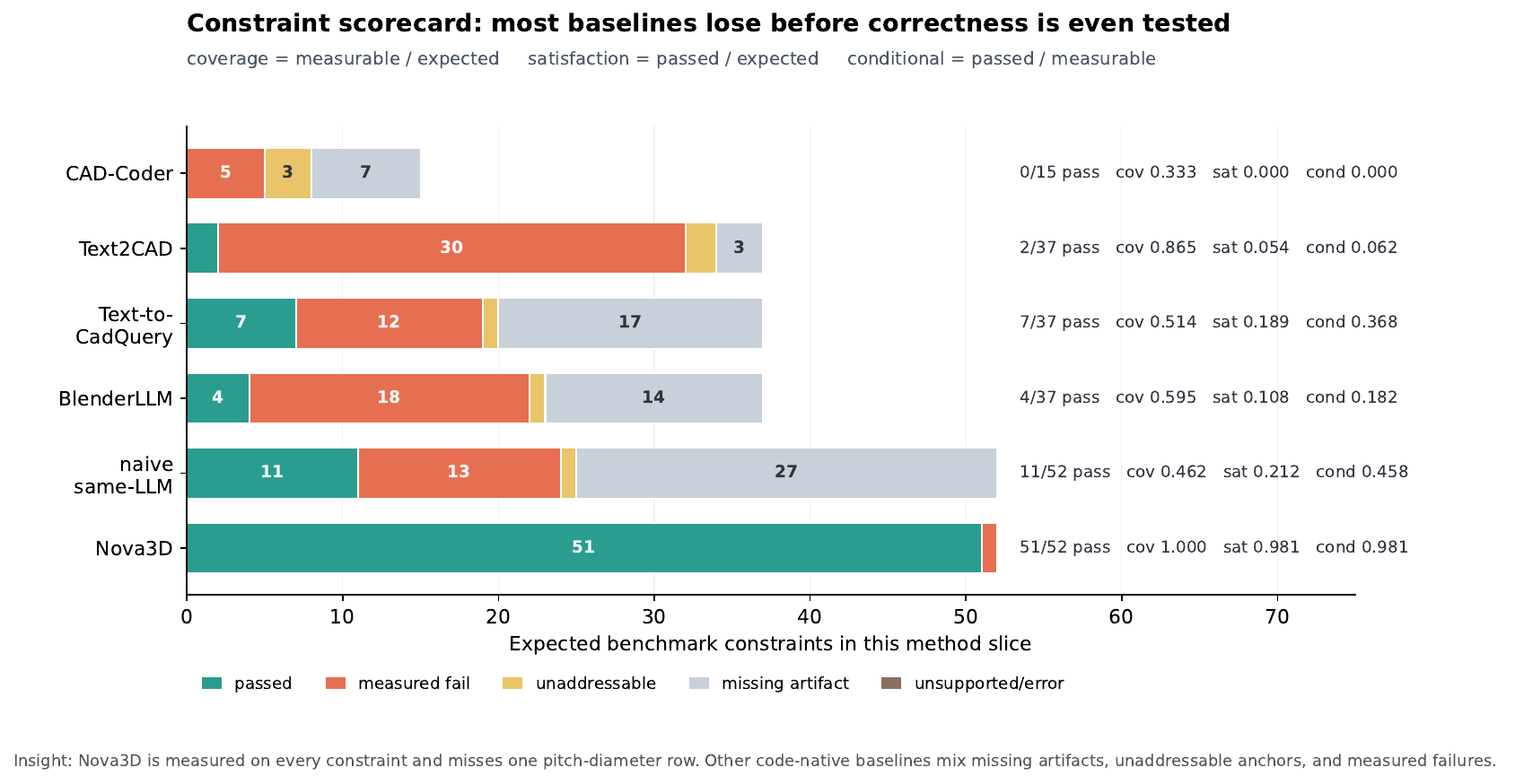}
\caption{\textbf{Constraint scorecard.} Stacked per-method bars: passed (green),
measured failures (orange), unaddressable anchors (yellow), and missing artifacts
(grey). Nova3D is measured on every row and misses one; baselines fail through
missing artifacts, missing anchors, or wrong values.}
\label{fig:constraints}
\end{figure}

Structure is only useful if it is measurable, not decorative. We test whether
assets satisfy the frozen numeric and count constraints in the specs: the scorer
loads the exported GLB, anchors semantic nodes from the scene graph, measures
counts or dimensions directly from geometry, and applies the declared comparator
and tolerance. We report \emph{coverage} (fraction measurable), \emph{satisfaction}
(fraction passing), \emph{conditional accuracy} (pass rate among measurable), and
the count-constraint pass rate. Nova3D is the only method with full formal coverage
and passes \textbf{51 of 52 constraints (98.1\%)}, including \textbf{32/32 count
constraints}; the best baseline passes 11/52 and all CAD/text-CAD baselines pass
\emph{zero} counts (Table~\ref{tab:constraints}, Figure~\ref{fig:constraints}). The count split is the strongest
structural proof: counting ``four wheels'' or ``twelve teeth'' requires the parts to
exist as separable named entities. The single Nova3D miss is a small gear's pitch
diameter (0.0695\,m vs 0.06\,m target at 10\% tolerance), a pitch-circle-vs-envelope
issue left as a fail.

\begin{table}[ht!]
\centering
\small
\caption{\textbf{Constraint satisfaction, measured from the exported GLB.} Item sets
differ by native modality. Satisfaction $=$ coverage $\times$ conditional accuracy.
Nova3D is the only method that is both measurable and correct on essentially every
row.}
\label{tab:constraints}
\begin{tabular}{lrrrrrrr}
\toprule
Method & Constraints & Measurable & Passed & Coverage & Satisfaction & Conditional & Counts \\
\midrule
CAD-Coder & 15 & 5 & 0 & 0.333 & 0.000 & 0.000 & 0/10 \\
Text2CAD & 37 & 32 & 2 & 0.865 & 0.054 & 0.062 & 0/22 \\
Text-to-CadQuery & 37 & 19 & 7 & 0.514 & 0.189 & 0.368 & 0/22 \\
BlenderLLM & 37 & 22 & 4 & 0.595 & 0.108 & 0.182 & 0/22 \\
naive same-LLM & 52 & 24 & 11 & 0.462 & 0.212 & 0.458 & 6/32 \\
\textbf{Nova3D} & \textbf{52} & \textbf{52} & \textbf{51} & \textbf{1.000} & \textbf{0.981} & \textbf{0.981} & \textbf{32/32} \\
\bottomrule
\end{tabular}
\end{table}

\paragraph{Worked example.}
The aggregate is strong, but a worked measurement shows the metric is an audit
chain, not magic. For the city bicycle, the scorer anchors semantic nodes in the
exported scene graph (the two wheel sub-assemblies, the front basket) and
measures directly from geometry (Table~\ref{tab:bike}). Each row reduces the score
to a concrete anchor, a measured value, and a tolerance verdict.

\begin{table}[ht!]
\centering
\caption{\textbf{Worked constraint audit (city bicycle).} Measurements are taken
from named anchors in the exported GLB and compared to the frozen spec.}
\label{tab:bike}
\begin{tabular}{lrrl}
\toprule
Constraint & Target & Measured & Verdict \\
\midrule
wheel\_count & 2 & 2 & pass \\
wheel\_diameter & 0.66\,m $\pm$10\% & 0.659\,m & pass \\
fender\_count & 2 & 2 & pass \\
basket\_count & 1 & 1 & pass \\
\bottomrule
\end{tabular}
\end{table}

\section{Local Editability}
\label{sec:editability}
\begin{figure}[ht!]
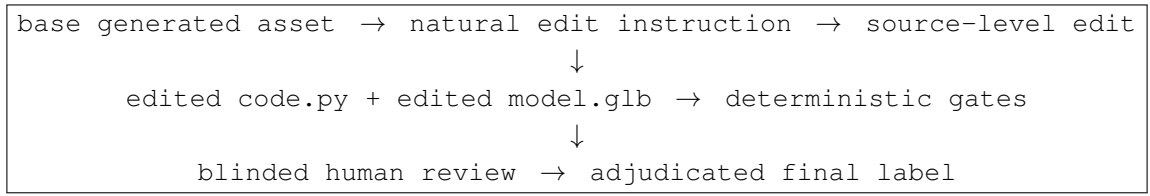

\centering
\fbox{\begin{minipage}{0.9\linewidth}\ttfamily\small
\centering
base generated asset \;$\rightarrow$\; natural edit instruction \;$\rightarrow$\; source-level edit\\[2pt]
$\downarrow$\\[2pt]
edited code.py + edited model.glb \;$\rightarrow$\; deterministic gates\\[2pt]
$\downarrow$\\[2pt]
blinded human review \;$\rightarrow$\; adjudicated final label
\end{minipage}}
\caption{\textbf{Local-edit protocol.} Deterministic representation gates precede a
blinded, adjudicated human visual review.}
\label{fig:editprotocol}
\end{figure}

Programmable structure should make \emph{local} editing possible, which is the
bridge to interactive world-building. We evaluate 18 generated assets (six
domains, three levels), each receiving one neutral edit instruction (13
additive, e.g. ``add a headrest''; 5 modifications, e.g. ``make the wheels
chunkier''), through the protocol in Figure~\ref{fig:editprotocol}. A
deterministic core checks artifact validity, target addressability, source/GLB
change, and hierarchy preservation; visual target semantics and locality are
then scored by two human reviewers with adjudication.

The reviewers are project-team members; consistent with the benchmark's stated
no-independent-annotation policy (Section~\ref{sec:bench}), we do not claim
external raters,
and instead report chance-corrected agreement and release the review panels.
Reviews are blinded in the sense that each reviewer scores randomized base/edit
render panels that hide the deterministic gate outcomes, run metadata, and the
other reviewer's labels; only the edit instruction and the two panels are shown.

\begin{figure}[ht!]
\centering
\begin{subfigure}{0.46\linewidth}
\includegraphics[width=\linewidth]{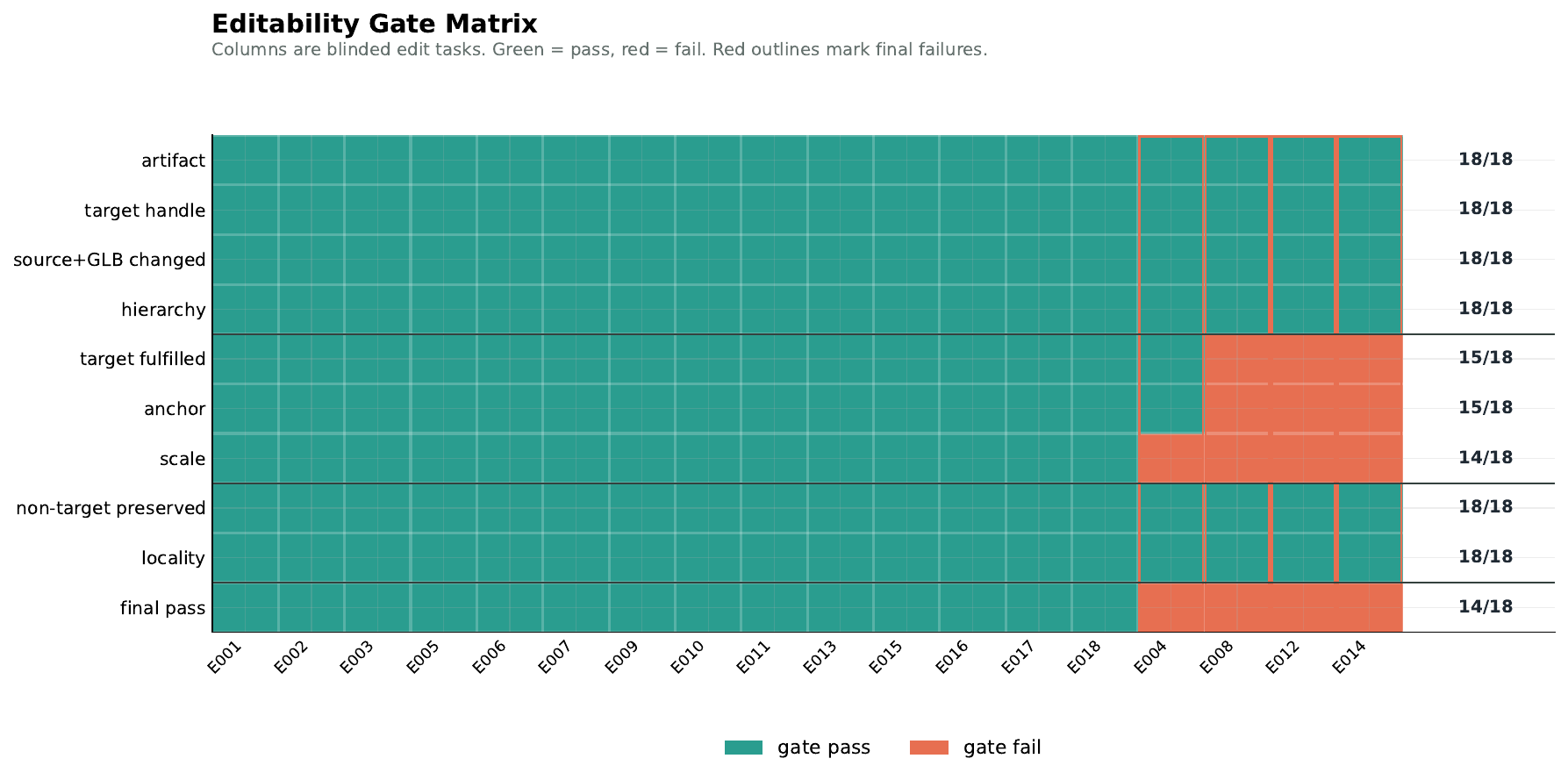}
\caption{Gate matrix}
\end{subfigure}\hfill
\begin{subfigure}{0.5\linewidth}
\includegraphics[width=\linewidth]{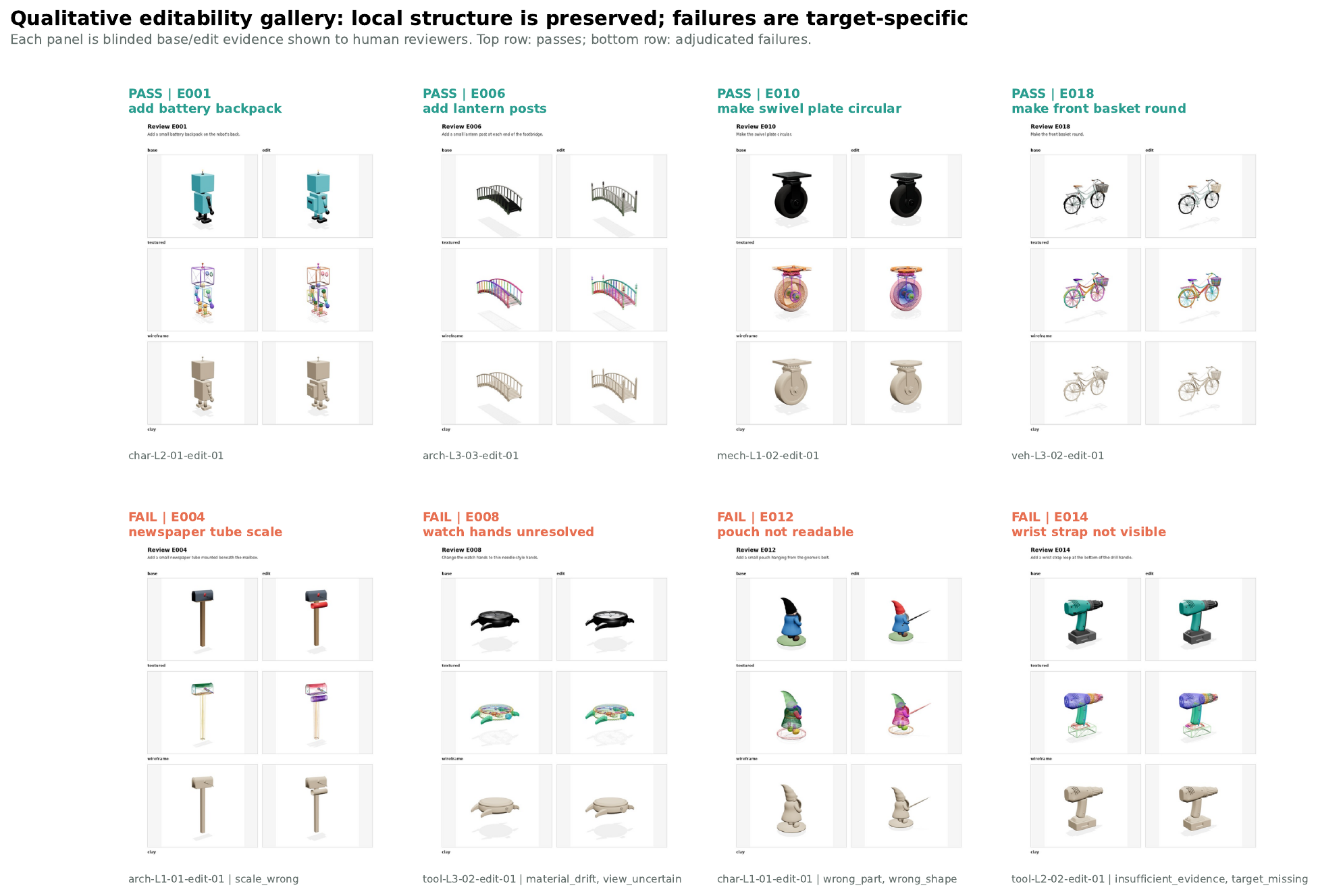}
\caption{Qualitative gallery}
\end{subfigure}
\caption{\textbf{Local edits preserve the surrounding asset.} Deterministic
representation gates saturate at 18/18; the limiting stage is target semantics. In
both successful and failed edits the surrounding asset stays intact. Failures are
local targets that are not visually correct enough, not global regenerations.}
\label{fig:editability}
\end{figure}

The deterministic gates saturate at \textbf{18/18}, and the final adjudicated
result is \textbf{14/18 successful local edits} (additive 10/13, modifications
4/5; Table~\ref{tab:editability}, Figure~\ref{fig:editability}). Critically, both reviewers agreed that
\textbf{all 18/18} edits preserved non-target content and passed the locality
gate: the four failures are target-semantic (a too-small newspaper tube,
unresolved watch hands, an unrecognizable belt pouch, a missing wrist strap),
not global rewrites. Chance-corrected agreement is perfect exactly on the
locality claims ($\kappa = 1.0$ for non-target preservation and locality) and
moderate on target fulfillment ($\kappa = 0.46$), which is the field the
adjudication step exists to resolve (Table~\ref{tab:editagreement}). The
supported claim is precise: Nova3D makes local programmatic edits while
preserving the rest of the asset. This is not perfect editing and not a baseline
leaderboard, and target-level precision for small accessories is the limiting
factor.

\begin{table}[ht!]
\centering
\begin{minipage}[t]{0.46\linewidth}
\centering
\caption{\textbf{Local editability outcomes.} Deterministic gates saturate; the final
adjudicated visual pass is 14/18, with non-target preservation and locality 18/18.}
\label{tab:editability}
\begin{tabular}{lrr}
\toprule
Slice & N & Final pass \\
\midrule
All edits & 18 & 14 (77.8\%) \\
Additive & 13 & 10 (76.9\%) \\
Modified-existing & 5 & 4 (80.0\%) \\
Text-conditioned base & 13 & 11 (84.6\%) \\
Image-conditioned base & 5 & 3 (60.0\%) \\
\midrule
Non-target preserved & 18 & 18 (100\%) \\
Locality preserved & 18 & 18 (100\%) \\
\bottomrule
\end{tabular}
\end{minipage}\hfill
\begin{minipage}[t]{0.5\linewidth}
\centering
\caption{\textbf{Reviewer agreement} (before adjudication). Chance-corrected
agreement is perfect on the locality claims and moderate on target fulfillment.}
\label{tab:editagreement}
\begin{tabular}{lrr}
\toprule
Field & Agreement & Cohen's $\kappa$ \\
\midrule
target fulfilled & 88.9\% & 0.455 \\
target anchor ok & 94.4\% & 0.769 \\
target scale ok & 94.4\% & 0.824 \\
non-target preserved & 100.0\% & 1.000 \\
locality pass & 100.0\% & 1.000 \\
overall pass & 94.4\% & 0.769 \\
\bottomrule
\end{tabular}
\end{minipage}
\end{table}

\section{Articulation and Interaction Readiness}
\label{sec:articulation}
\begin{figure}[ht!]
\centering
\begin{subfigure}{0.52\linewidth}
\includegraphics[width=\linewidth]{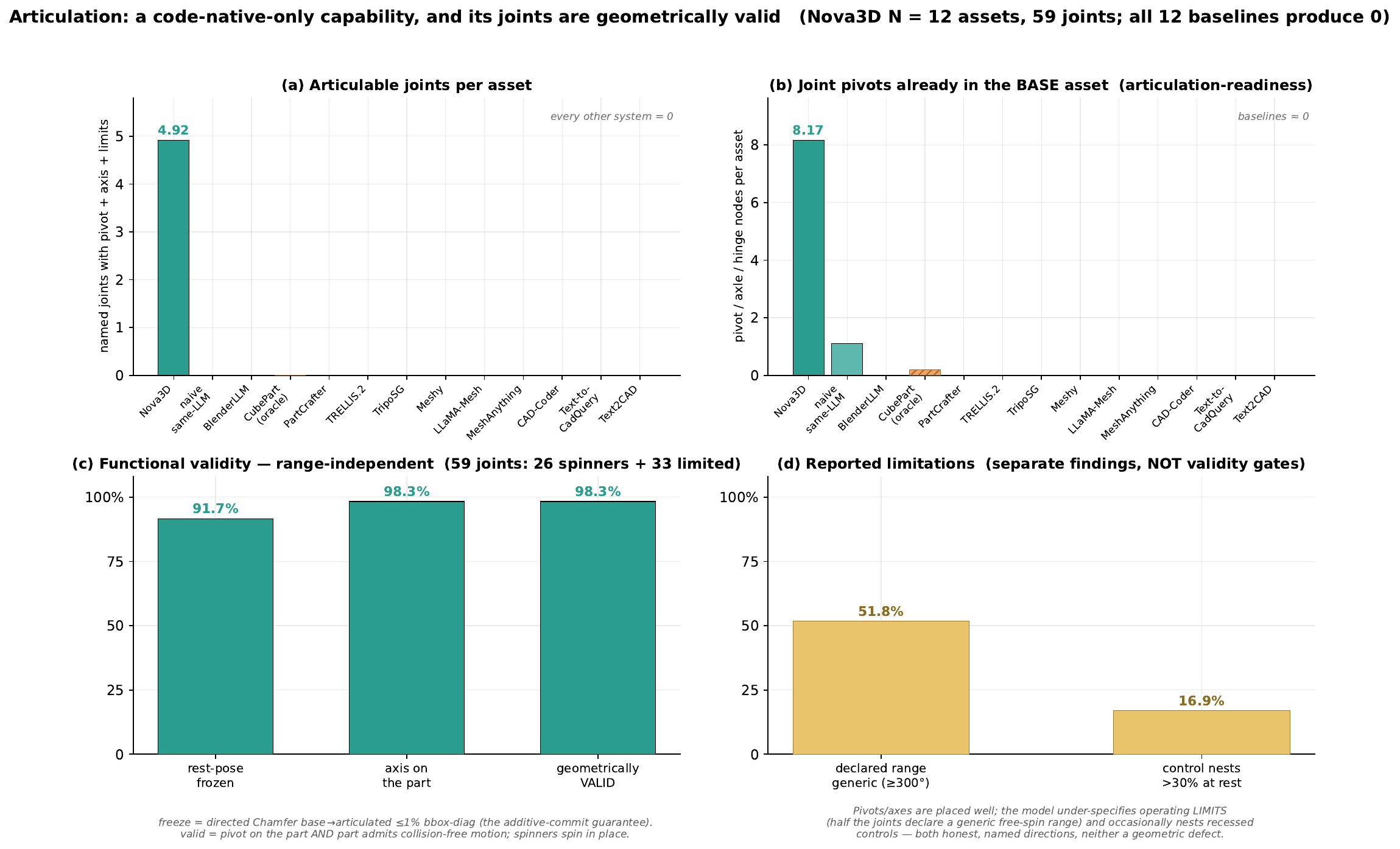}
\caption{Capability and validity}
\end{subfigure}\hfill
\begin{subfigure}{0.45\linewidth}
\includegraphics[width=\linewidth]{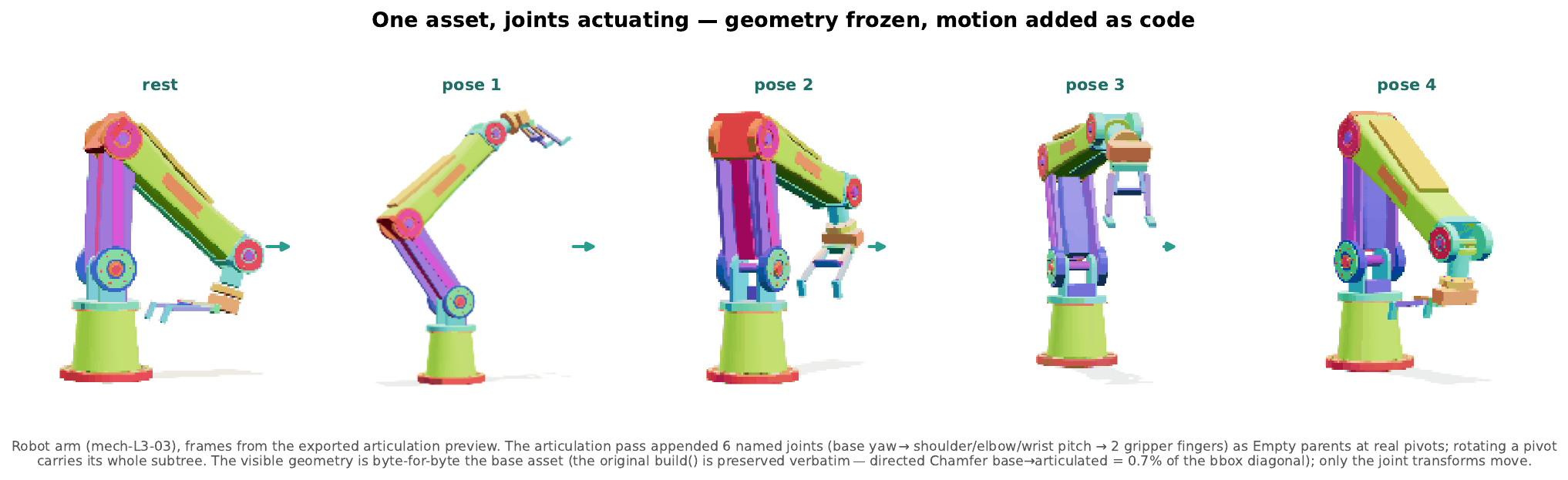}
\caption{Robot-arm motion (geometry frozen)}
\end{subfigure}
\caption{\textbf{Articulation as an additive operation on a program.} Nova3D produces
articulable joints and readiness pivots while all baselines are at zero; the robot
arm actuates across poses while its visible geometry stays frozen. Motion is added
as code, not regenerated.}
\label{fig:articulation}
\end{figure}

The final downstream operation is turning a static asset into a moving one. A
joint must attach to a part, so articulation is available natively only when
parts exist as addressable entities. Across the full corpus, Nova3D assets
average \textbf{8.17 readiness pivots per asset} (measured in the hierarchy
axis); on a stratified case study of 12 articulated assets (59 joints), Nova3D
produces \textbf{4.92 joints per asset} while every mesh-native, CAD, and
segmentation baseline produces \textbf{zero}
(Table~\ref{tab:articulation}, Figure~\ref{fig:articulation}). Of the joints an object should have, Nova3D
assigns the correct kinematic class with accuracy \textbf{0.955} and recovers
them with recall \textbf{0.761}, and \textbf{58/59 (98.3\%)} joints are
geometrically valid (axis on the moving part, and the part admits
collision-free motion), with the original geometry frozen under articulation on
11/12 assets (Table~\ref{tab:articvalidity}). The one disclosed failure is a
pendulum sign whose axis sits on its support bar. This is a capability case
study, not a large statistical benchmark. The strong claim is categorical:
baselines expose no native joints. The careful claim is that Nova3D's joints
are geometrically valid but their operating \emph{limits} are generic (29/56
revolute joints declare $\ge 300^\circ$) and need future work.

\begin{table}[ht!]
\centering
\begin{minipage}[t]{0.5\linewidth}
\centering
\caption{\textbf{Articulation capability} (Tier 1). No baseline exposes native joints.}
\label{tab:articulation}
\begin{tabular}{lcr}
\toprule
Method & Articulable? & Joints/asset \\
\midrule
\textbf{Nova3D} & yes & \textbf{4.92} \\
naive same-LLM & no & 0 \\
CubePart (oracle) & no & 0 \\
mesh-native / CAD / seg. & no & 0 \\
\bottomrule
\end{tabular}
\end{minipage}\hfill
\begin{minipage}[t]{0.46\linewidth}
\centering
\caption{\textbf{Spec correctness \& functional validity} (Nova3D, 12 assets).}
\label{tab:articvalidity}
\begin{tabular}{lr}
\toprule
Metric & Value \\
\midrule
Articulated assets & 12 \\
Total joints & 59 \\
Revolute / prismatic & 56 / 3 \\
Type accuracy & 0.955 \\
Joint recall & 0.761 \\
\;\; required / optional & 0.700 / 0.722 \\
Extra joints beyond spec & 29 \\
Rest pose frozen & 11 / 12 \\
Axis on moving part & 58 / 59 \\
Geometrically valid & 58 / 59 \\
Generic-range issue & 29 / 56 rev. \\
\bottomrule
\end{tabular}
\end{minipage}
\end{table}

\section{Production Readiness from Source Representation}
\begin{figure}[ht!]
\centering
\begin{subfigure}{0.5\linewidth}
\includegraphics[width=\linewidth]{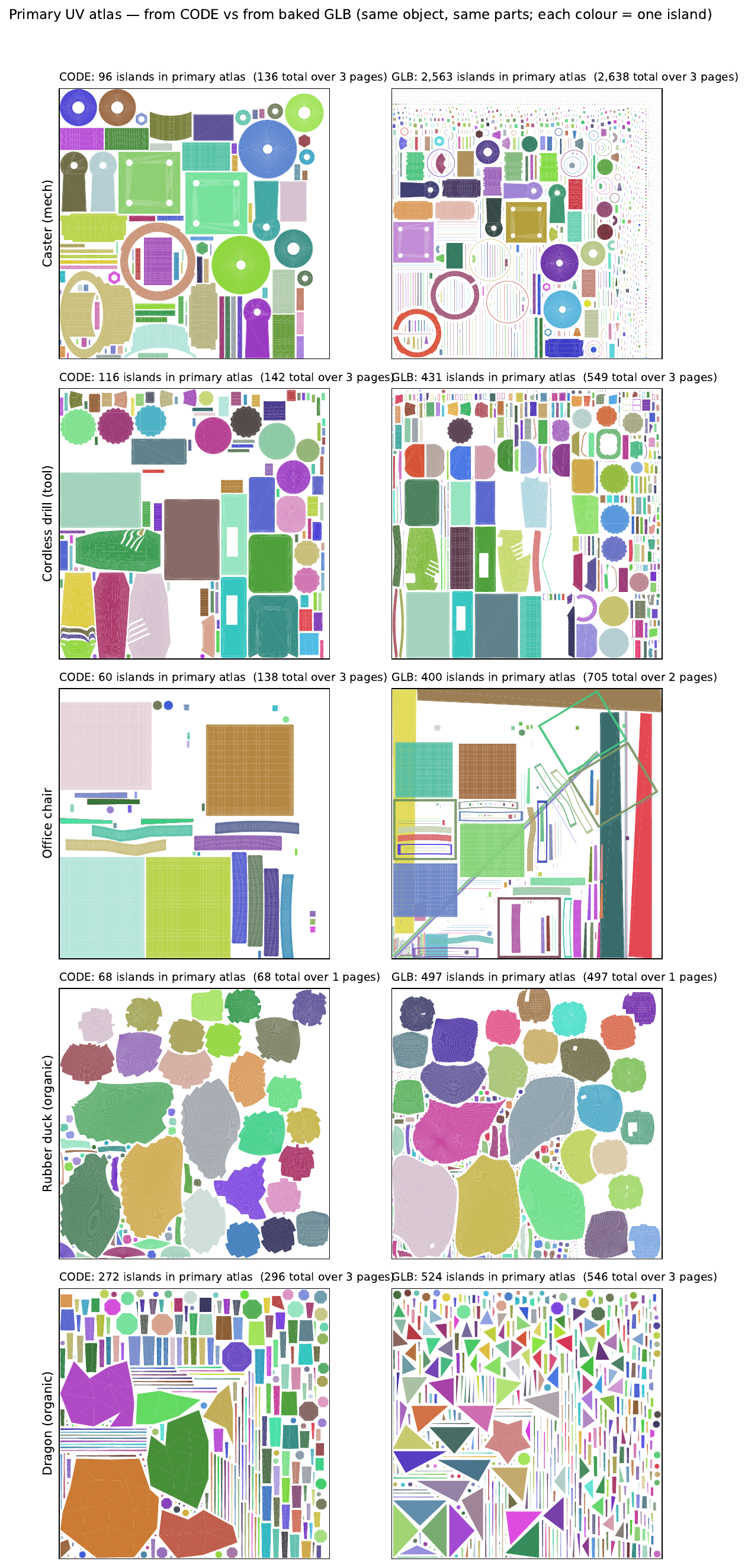}
\caption{UV atlases: code vs baked GLB}
\end{subfigure}\hfill
\begin{subfigure}{0.46\linewidth}
\includegraphics[width=\linewidth]{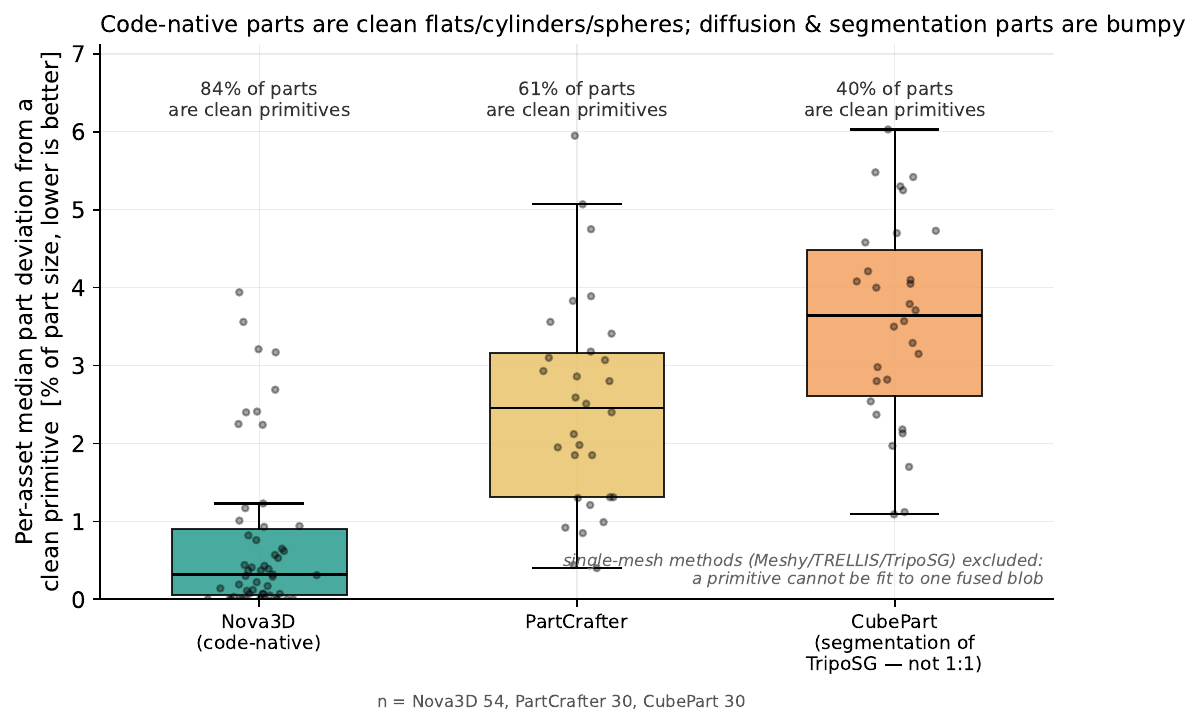}
\caption{Per-part primitive residual}
\end{subfigure}
\caption{\textbf{Production readiness from source representation.} (a) Unwrapping
the pre-modifier construction geometry yields few large UV islands; unwrapping the
baked GLB shatters the surface into many small islands. (b) Code-built parts
deviate far less from clean fitted primitives than diffusion- or
segmentation-derived parts.}
\label{fig:production}
\end{figure}

Keeping the asset as source code preserves production affordances that are damaged
when the same asset is flattened to a baked mesh. These are not random extras but
direct consequences of the representation (Table~\ref{tab:affordances},
Figure~\ref{fig:production}).

\begin{table}[ht!]
\centering
\caption{\textbf{Source-derived production affordances.} Each affordance follows
from keeping the asset as clean, part-structured source code.}
\label{tab:affordances}
\begin{tabular}{L{3.0cm} L{4.3cm} L{5.2cm}}
\toprule
Affordance & Why code helps & Evidence \\
\midrule
Clean primitive surfaces & parts are built from explicit primitives and modifiers & 0.3\% median primitive residual; 84\% primitive-clean \\
UV readiness & the pre-modifier construction geometry can be unwrapped before mesh baking & 214 vs 1{,}345 median islands (6.0$\times$ fewer); code wins 54/54 \\
Material intent & semantic parts receive material families by meaning & PBR-style material parameters, emissive flags, vertex AO \\
Compact source & the asset is stored as source, not only binary geometry & 18.7\,KB source program \\
\bottomrule
\end{tabular}
\end{table}

\textbf{Surface finish.} Fitting the best of \{plane, cylinder, sphere\} to each
part, a Nova3D part deviates by a median of \textbf{0.3\%} of its size (84\%
primitive-clean), versus 2.5\% for PartCrafter and 3.6\% for CubePart, whose
segmentation-derived parts carry TripoSG's surface plus the cut. The diffusion-
and segmentation-derived parts are $6$--$11\times$ bumpier. This is a per-part
metric, so single-mesh methods cannot enter it.

\textbf{UV-readiness.} In a controlled paired ablation on all 54 items (same
object, same named parts, same unwrapper, changing only the representation),
unwrapping the pre-modifier construction geometry, the base cage, yields a
median of \textbf{214 UV islands} versus \textbf{1{,}345} from the compiled GLB,
a \textbf{6$\times$} reduction, and the code wins all 54 items (Wilcoxon
$p=1.6\times10^{-10}$; Table~\ref{tab:uv}). Because the baked condition uses our
own clean, still-part-structured geometry, it is a charitable stand-in for the
mesh-native setting; we report this as a representation claim (``code unwraps
cleaner than the mesh it compiles to''), not a head-to-head against a specific
system's delivered UVs.

\textbf{Engine-ready materials.} The program emits material \emph{intent}
directly: PBR-style material parameters assigned by part meaning, emissive flags
on light sources, and ambient occlusion baked into vertex colors with no UVs and
no texture files. Despite shipping no texture maps, this lightweight approach is
even with TripoSG on texture quality and well ahead of PartCrafter, doing the
first hour of a 3D artist's job automatically (implementation detail in the
appendix).

\begin{table}[ht!]
\centering
\caption{\textbf{UV-readiness, code vs compiled mesh} (paired, N=54). Median UV
island count per atlas; lower is better. Code wins all 54 items.}
\label{tab:uv}
\begin{tabular}{lrrr}
\toprule
Domain & Code islands & GLB islands & Ratio (GLB $\div$ code) \\
\midrule
Architecture & 347 & 2{,}677 & 13.1$\times$ \\
Tools & 233 & 2{,}199 & 9.4$\times$ \\
Mechanical & 267 & 2{,}021 & 6.5$\times$ \\
Furniture & 138 & 705 & 5.5$\times$ \\
Vehicles & 576 & 3{,}083 & 5.3$\times$ \\
Characters (organic) & 168 & 624 & 4.0$\times$ \\
\textbf{All (median)} & \textbf{214} & \textbf{1{,}345} & \textbf{6.0$\times$} \\
\bottomrule
\end{tabular}
\end{table}

\section{Discussion}
The results form one argument rather than eleven. Nova3D reliably generates
executable 3D code where the same base model without the system package does not;
the resulting assets are visually competitive; the code representation exposes
native named parts and a real assembly tree; and that structure is what makes
constraints measurable, edits local, and articulation possible. Read together,
these are not separate wins but a chain in which each link depends on the previous
one.

Code-native and mesh-native generation are complementary. Mesh-native methods
remain stronger for texture realism and freeform organic forms, and our perceptual
results place Nova3D close behind the best of them rather than ahead, which is the
right framing for a representation paper. Where Nova3D leads is on programmable
affordances: handles, hierarchy, constraints, edits, pivots, and articulation.
Post-hoc segmentation is not an equivalent substitute: it requires a finished
mesh, is often flat, may need an oracle vocabulary, and does not recover the
source construction logic. The naive same-LLM ablation shows that these
affordances come from the Nova3D system package, not from an LLM writing code in
general.

\begin{quote}
\emph{Nova3D's advantage is not that code is a prettier surface generator; it is
that code preserves the asset as an object a downstream system can inspect,
measure, modify, and animate.}
\end{quote}

\section{Limitations}
We state limitations explicitly rather than folding them into the discussion.
Nova3D-Bench has $N=54$ items, uses synthetic reference images, and relies on
AI-assisted, author-adjudicated annotation with no independent human spec
annotation. The perceptual VLM judge follows GPTEval3D but is not yet
human-validated on normal-render inputs specifically, and sees only two views.
The editability and articulation results are stratified case studies (18 and 12
assets), not statistical samples; the editability reviewers are project-team
members rather than external raters; and articulation operating limits are
generic LLM defaults rather than physically validated ranges. Geometric
integrity is shared with part-aware diffusion (``cleanest among part-aware
methods,'' not unique). Hierarchy is scored for existence, depth, and pivots,
not provable nesting correctness. Texture realism is conceded to baked-PBR
systems, and organic characters remain weaker. Finally, the representation
depends on Blender/Python as the execution substrate.

\section{Conclusion}
This paper reframes 3D generation from surface synthesis to source-code
synthesis. By generating an executable program whose compiled mesh is merely the
artifact, Nova3D produces assets that are visually competitive while natively
exposing the structure interactive worlds require: named parts, an assembly
hierarchy, pivots, and edit handles. Constraints can therefore be measured
against the source's named anchors, edits stay local, and articulation is an
additive operation on the program. Across reliability, visual quality,
structure, constraints, editing, and articulation, the evidence supports a
single claim: code-native generation turns a generated 3D object from an opaque
surface into a programmable asset. Programmable assets are the bridge from
one-off renders toward interactive 3D worlds.



\begin{thebibliography}{99}

\bibitem{poole2022dreamfusion}
Ben Poole et al. \textit{DreamFusion: Text-to-3D using 2D Diffusion}. 2022.

\bibitem{lin2023magic3d}
Chen-Hsuan Lin et al. \textit{Magic3D: High-Resolution Text-to-3D Content
Creation}. 2023.

\bibitem{nichol2022pointe}
Alex Nichol et al. \textit{Point-E: A System for Generating 3D Point Clouds from
Complex Prompts}. 2022.

\bibitem{jun2023shape}
Heewoo Jun and Alex Nichol. \textit{Shap-E: Generating Conditional 3D Implicit
Functions}. 2023.

\bibitem{liu2023zero123}
Ruoshi Liu et al. \textit{Zero-1-to-3: Zero-shot One Image to 3D Object}. 2023.

\bibitem{li2025triposg}
Yangguang Li et al. \textit{TripoSG: High-Fidelity 3D Shape Synthesis using
Large-Scale Rectified Flow Models}. 2025.

\bibitem{xiang2024trellis}
Jianfeng Xiang et al. \textit{Structured 3D Latents for Scalable and Versatile 3D
Generation (TRELLIS)}. 2024.

\bibitem{zhao2025hunyuan3d}
Tencent Hunyuan3D Team. \textit{Hunyuan3D 2.0: Scaling Diffusion Models for
High-Resolution Textured 3D Asset Generation}. 2025.

\bibitem{meshy2024}
Meshy. \textit{Meshy 5: Commercial Text/Image-to-3D Generation}. 2024.

\bibitem{mo2019partnet}
Kaichun Mo et al. \textit{PartNet: A Large-Scale Benchmark for Fine-Grained and
Hierarchical Part-Level 3D Object Understanding}. 2019.

\bibitem{liu2023partslip}
Minghua Liu et al. \textit{PartSLIP: Low-Shot Part Segmentation for 3D Point Clouds
via Pretrained Image-Language Models}. 2023.

\bibitem{yang2024sampart3d}
Yunhan Yang et al. \textit{SAMPart3D: Segment Any Part in 3D Objects}. 2024.

\bibitem{ma2024find3d}
Ziqi Ma et al. \textit{Find Any Part in 3D (Find3D)}. 2024.

\bibitem{liu2024part123}
Anran Liu et al. \textit{Part123: Part-aware 3D Reconstruction from a Single-view
Image}. 2024.

\bibitem{chen2024partgen}
Minghao Chen et al. \textit{PartGen: Part-level 3D Generation and Reconstruction
with Multi-view Diffusion Models}. 2024.

\bibitem{lin2025partcrafter}
Yuchen Lin et al. \textit{PartCrafter: Structured 3D Mesh Generation via
Compositional Latent Diffusion Transformers}. 2025.

\bibitem{paul2025namethatpart}
Sumit Paul et al. \textit{Name That Part: 3D Part Segmentation and Naming
(ALIGN-Parts)}. 2025. arXiv:2512.18003.

\bibitem{sun20243dgpt}
Chunyi Sun et al. \textit{3D-GPT: Procedural 3D Modeling with Large Language
Models}. 2024.

\bibitem{yamada2024l3go}
Yutaro Yamada et al. \textit{L3GO: Language Agents with Chain-of-3D-Thoughts for
Generating Unconventional Objects}. 2024.

\bibitem{hu2024scenecraft}
Ziniu Hu et al. \textit{SceneCraft: An LLM Agent for Synthesizing 3D Scenes as
Blender Code}. 2024.

\bibitem{ll3m2024}
Sining Lu, Guan Chen, Nam Anh Dinh, Itai Lang, Ari Holtzman, and Rana Hanocka.
\textit{LL3M: Large Language 3D Modelers}. 2025. arXiv:2508.08228.

\bibitem{du2024blenderllm}
Yuhao Du et al. \textit{BlenderLLM: Training Large Language Models for
Computer-Aided Design with Self-improvement}. 2024.

\bibitem{wu2021deepcad}
Rundi Wu et al. \textit{DeepCAD: A Deep Generative Network for Computer-Aided
Design Models}. 2021.

\bibitem{khan2024text2cad}
Mohammad Sadil Khan et al. \textit{Text2CAD: Generating Sequential CAD Models from
Beginner-to-Expert Level Text Prompts}. 2024.

\bibitem{texttocadquery2025}
Haoyang Xie and Feng Ju. \textit{Text-to-CadQuery: A New Paradigm for CAD
Generation with Scalable Large Model Capabilities}. 2025. arXiv:2505.06507.

\bibitem{cadcoder2025}
Anna C. Doris, Md Ferdous Alam, Amin Heyrani Nobari, and Faez Ahmed.
\textit{CAD-Coder: An Open-Source Vision-Language Model for Computer-Aided
Design Code Generation}. 2025. arXiv:2505.14646.

\bibitem{cadfusion2025}
Ruiyu Wang, Yu Yuan, Shizhao Sun, and Jiang Bian. \textit{Text-to-CAD
Generation Through Infusing Visual Feedback in Large Language Models
(CADFusion)}. ICML 2025. arXiv:2501.19054.

\bibitem{barkley2026cadsmith}
J. Barkley et al. \textit{CADSmith: Multi-Agent CAD Generation with
Programmatic Geometric Validation}. 2026. arXiv:2603.26512.

\bibitem{zhang2026cadtestbench}
Anonymous. \textit{Text-to-CAD Evaluation with CADTests (CADTestBench)}.
2026. arXiv:2605.07807.

\bibitem{wu2024gpteval3d}
Tong Wu et al. \textit{GPT-4V(ision) is a Human-Aligned Evaluator for Text-to-3D
Generation}. CVPR 2024.

\bibitem{he2023t3bench}
Yuze He et al. \textit{T3Bench: Benchmarking Current Progress in Text-to-3D
Generation}. 2023.

\bibitem{su2024gt23dbench}
Sitong Su et al. \textit{GT23D-Bench: A Comprehensive General Text-to-3D Generation
Benchmark}. 2024.

\bibitem{deitke2023objaverse}
Matt Deitke et al. \textit{Objaverse: A Universe of Annotated 3D Objects}. 2023.

\bibitem{koch2019abc}
Sebastian Koch et al. \textit{ABC: A Big CAD Model Dataset for Geometric Deep
Learning}. 2019.

\bibitem{willis2021fusion360}
Karl D.D. Willis et al. \textit{Fusion 360 Gallery: A Dataset and Environment for
Programmatic CAD Construction from Human Design Sequences}. 2021.

\bibitem{li2025gencadcode}
Anna C. Doris, Md Ferdous Alam, Amin Heyrani Nobari, and Faez Ahmed.
\textit{GenCAD-Code: A Dataset of CAD-Model Image and Code Pairs} (introduced
with CAD-Coder). 2025. arXiv:2505.14646.

\bibitem{wang2024llamamesh}
Zhengyi Wang et al. \textit{LLaMA-Mesh: Unifying 3D Mesh Generation with Language
Models}. 2024.

\bibitem{chen2024meshanything}
Yiwen Chen et al. \textit{MeshAnything: Artist-Created Mesh Generation with
Autoregressive Transformers}. 2024.

\end{thebibliography}
\end{document}